\documentclass{aa}  
\usepackage{graphicx}
\usepackage{color}

\usepackage{txfonts}
\usepackage{hyperref}
\def\h2{\hskip-2pt}

\begin{document}

   \titlerunning{Model fitting by phase matching .}

   \title{Surface layer independent model fitting by phase matching:\\  theory and application to HD49933 and HD177153 (aka Perky)}

   \author{Ian W. Roxburgh}

   \institute{Astronomy Unit, Queen Mary University of London, 
     Mile End Road, London E1 4NS, UK.
   \email {I.W.Roxburgh@qmul.ac.uk} }

   \date{Received  / Accepted  }

 
   \abstract
    {}
{To describe the theory of surface layer independent model fitting by phase matching  and to apply this to the stars HD49933 observed by CoRoT, and HD177153 (aka Perky), 
observed by Kepler}
{We use theoretical analysis, phase shifts, and model fitting. }
{We define the inner and outer phase shifts of a frequency set of a model star and show that the outer phase shifts are (almost) independent of degree $\ell$, and that a function of the inner phase shifts (the phase function) collapses to an $\ell$ independent function of frequency in the outer layers. 
We then show how to use this result in a model fitting technique to find a best fit model to an observed frequency set by calculating the inner phase shifts of a model using the
 \emph{observed} frequencies and determining the extent to which the phase function  collapses to a single function of frequency in the outer layers.  We give two examples applying this technique to the frequency sets of HD49933 observed by CoRoT  and HD177153 (aka Perky) observed by Kepler, and compare our results with those of previous studies and show that they are compatible with those obtained using different techniques.  We show that there can be many different models that fit the data within the errors and that  better precision on the frequencies is needed to discriminate between the models.  We compare this technique to that using the ratios of small to large separations, showing that in principle it is more accurate and avoids the problem of correlated errors in separation ratio  fitting.} 
    {}
   \keywords{stars: oscillations, - asteroseismology - stars:  interiors - methods: analytical - methods: numerical }

   \maketitle
\section{Introduction}

At first sight, the obvious way to seek to infer the internal structure of a star with  an observed frequency  set $\nu^{obs}_{n\ell}$, with error estimated $\sigma_{n\ell}$  
is to search for a model (or models)  whose frequencies $\nu^{mod}_{n\ell}$ fit the observed values  within the error estimates.  However it has long been appreciated that this is not so straight forward: modelling of the outer layers of a star is subject to considerable uncertainties (cf Christensen-Dalsgaard et al, 1988, Dziembowski et al 1988)
due to our poor understanding of the physical processes that determine the structure of these layers; these include
 modelling convection, convective 
overshooting, non-adiabatic effects on both convection and oscillations, turbulent pressure, the equation of state, diffusion, mild turbulence, magnetic fields, rotation and global circulation. All these factors impact on the oscillation frequencies of a model star, and therefore hinder efforts to find stellar models whose frequencies best fit an observed frequency set.  Even for the solar case observations and models disagree by up to $\sim 10\mu$Hz. 

One way of seeking to overcome these problems is the ``frequency offset technique" (Kjeldsen et al 2008), in which the difference between the observed solar frequencies and those of a ``best solar model" is fitted by a power law $a\,\nu^b$, which is then scaled by a single factor (determined by the average frequency and large separation) and applied to other stars when seeking a best fit model.  This technique is 
widely used and has been incorporated into the Asteroseismic Modeling Portal (AMP) software  (Metcalfe et al 2009).  However the assumption that the
 many 
 differences in the properties of the outer layers of  
 different 
 stars  can be captured in a single scaling factor  remains to be verified. 

A second method is  fitting separation ratios: that is finding models whose frequencies are such that their ratios of small to large separations best fit the same ratios of the observed frequency set, since such ratios  subtract off the major effects of the outer layers (Roxburgh and Vorontsov 2003). But note that, as pointed out in Roxburgh and Vorontsov (2013), it is not 
the ratios at the same $n$ values that should be compared but rather the model values interpolated to the observed frequencies.  Note also that since  the ratios only depend on the structure of the inner layers, they can only give information on the interior of a star and not on the outer layers.

We here present an alternative surface layer independent procedure (Phase Matching) that rests on the result that in the outer layers of a star the phase shift $\alpha_\ell(\nu)$ of the eigenfunction of an oscillation mode (that is the departure from a pure harmonic function)  is almost independent of the angular degree $\ell$,  so that even though we do not know what $\alpha(\nu)$ is, for the model to fit the frequencies a function ${\cal G}(\ell,\nu)$ (the phase function) of the  inner phase shifts
 $\delta_\ell(\nu)$, calculated at the {\it observed} frequencies, must be such that it matches on to a function only of frequency in the outer layers. 

Details of the analysis are given in section 2, illustrated using models of a $1.15  M_\odot$ star in different stages of evolution, primarily  Model A (mid main sequence); Model B 
(terminal main sequence) and Model C (post main sequence in the shell burning phase). 
In section 3 this procedure is applied to the frequency sets of  the CoRoT star HD49933 and the  Kepler star  HD177153 (aka Perky) to find best fit models, and the results compared with those obtained by other authors: Piau et al (2009), Benomar et al (2010), Kallinger et al (2010), Bigot et al (2011), Creevey and Bazot (2011) and Liu et al (2014),  Silva-Aguire et al, (2013).  In section 4 we compare the method to that using separation ratios, showing that it is in principle more accurate and that it avoids  the problem of correlated errors when using separation ratios.  


\section{Theory of model fitting by phase matching}
\subsection{The phase shifts}
For any angular degree $\ell$ the frequencies of a model star are given as the eigenvalues of the 4th order system of equations  governing the oscillations, but for radial ($\ell=0$) modes  the 
 equations reduce to 2nd order in the variable $\psi_0(\omega,t)= r p' /(\rho c)^{1/2}$ 
$${d^2\psi_0\over dt^2}+Q_0 {d\psi_0\over dt}+\left( \omega^2-V_0\right) \psi_0 =0,~~~{\rm where}~~~ t=\int_0^r {dr\over c} \eqno(1)$$ 
is the acoustic radius, $p'$ the Eulerian pressure perturbation, $\omega=2\pi\nu$, and $Q_0(\omega\,t)$ and $V_0(\omega,\,t)$ are {\it acoustic potentials} that depend on both frequency and the structure  variables: density $\rho(r)$, pressure $P(r)$, sound speed $c(r)$ ($c^2=\Gamma_1 P/\rho$) and adiabatic exponent $\Gamma_1(r)$.  Were $Q_0$ and $V_0=0$, this is just the simple harmonic equation whose solution is  $\sin(\omega\, t)$.

Since Eqn 1 is  a 2nd order homogeneous equation it can be reduced to a  1st order inhomogeneous equation in the variable $\omega\psi_0/(d\psi_0/dt)$ and solved in terms of an
 {\it inner phase shift} $\delta_0(\omega\,t) $ defined by
$$ {\omega\psi_0\over d\psi_0/dt} =\tan[\omega\, t + \delta_0(\omega,\,t)] \eqno(2)$$
where $\delta_0(\omega\,t)$ satisfies the equation
$${d\delta_0\over dt}= -{V\over\omega}\,\sin[\omega\,t + \delta_0] +Q \sin(2[\omega\,t + \delta_0])   \eqno(3) $$
The  central boundary condition of regularity at $t=0$ requires  $\delta_0=0$ at $t=0$ \footnote{or $k\pi$ for integer $k$}; given the structure of the star $\delta_0(\omega,\,t)$ can be evaluated at any $t$ for any $\omega$ (not necessarily an eigenvalue),  and the value at any $t_f$ is determined solely by the structure interior to $t_f$. Of course in general Eqn 2 does not satisfy the surface boundary condition on $\psi_0$, the requirement that it do so determines the eigenfrequencies.

One can equally write Eqn 1 in terms of the acoustic depth $\tau=T-t$ where $T $ is the total acoustic radius of the star and represent the solution in the form
$${\omega\psi_0\over d\psi_0/d\tau} = \tan[\omega\, \tau - \alpha_0(\omega,\,\tau)] \eqno(4)$$
where $\alpha_0$ satisfies the surface boundary conditions on the gravitational potential and pressure perturbation \footnote{again subject to the addition of any multiple of $\pi$}. 
For $\ell=0$ modes the gravitational boundary condition that the potential and its derivative are  continuous with a solution of Laplace's equation is automatically satisfied (by conservation of mass and Newton's sphere theorem), and
$\alpha_0$ is a continuous function of $\omega$ and $\tau$,  determined solely by the structure of the outer layers above $\tau$. 
 
For an eigenfrequency $\omega\psi_0 / (d\psi_0/dt)$  must be continuous at all $t$, hence on equating the expressions in  Eqns  (2) and  (4) at any intermediate $t_f$,  and recalling that if $\tan A+\tan B =0$ then $A+B=n\pi$ for integer $n$, and that $\omega=2\pi\nu$, we obtain the result that an eigenfrequency $\nu_{n0}$ satisfies \footnote{this absorbs the arbitrary $k\pi$ in the boundary conditions}
$$ 2 \pi \nu_{n0} \,T  + \delta_0 (\nu_{n,0}\,t_f) - \alpha_0(\nu_{n0},\,\tau_f) = n\pi \eqno(5) $$
where $T=t+\tau=\int_0^R (dr/c) $ is the total acoustic radius.

For $\ell\ne 0$ we replace Eqns 2 and 4 by
$$ {2\pi\nu\psi_\ell\over d\psi_\ell/dt} = \tan[2\pi\nu t -\ell\pi/2+ \delta_{\ell}(\nu,t)]   \eqno(6a)$$
$$ {2\pi\nu\psi_\ell\over d\psi_\ell/d\tau} = \tan[2\pi\nu \tau-\alpha_{\ell}(\nu,\tau]   \eqno(6b)$$
Eqn 6a reflects the fact that the eigenfunctions behave like  spherical Bessel functions near $t=0$.
 Eqn 5 becomes 
$$2\pi T \nu_{n\ell} - \ell\pi/2   +  \delta_{n\ell} (\nu_{n\ell}, t_f) -\alpha_{n\ell} (\nu_{n\ell}, \tau_f)=     n\pi \eqno(7)$$
This is the {\it Eigenfrequency Equation} of Roxburgh and Vorontsov (2000, 2003).
Note that $n$ here is an unknown integer, it may be related to the radial order, as is the case for p-modes in main sequence stars, but it loses any such meaning in more evolved stars which have mixed modes. 

For $\ell=1$ the equations can again be reduced to 2nd order (Takata 2005, Roxburgh 2006, 2008a) so, for any $\nu$,  $\delta_1(\nu,\,t)$ is a continuous function of $t$ and $\alpha_1(\nu,\,\tau)$ a continuous function of $\tau$ (the surface condition on the gravitational potential being automatically satisfied by momentum conservation and MacCullagh's (1855) theorem).

For $\ell\ge 2$ the equations remain of 4th order. One can readily determine $\delta_\ell(\nu_{n\ell},\,t)$  and $\alpha_\ell(\nu_{n\ell},\,\tau)$ from Eqns 6a and  6b 
 for eigenfrequencies $\nu_{n\ell}$, but 
one can also determine $\delta_\ell(\nu,\,t)$ for any $\nu$, applying the gravitational boundary condition, but not the pressure condition, at the surface or anywhere in the outer layers where the density is negligible.
One can also determine a continuous $\alpha_\ell(\nu,\,\tau)$  by the (very good) approximation of  setting the oscillating gravitational potential and its derivative  to zero at the surface. (These approximations are discussed in the Appendix)
Alternatively for main sequence stars one can reduce the 4th order system of equations to second order by using the short wavelength Born approximation (Roxburgh and Vorontsov 1994a). 

The conclusion of this section is that given a stellar model we can determine continuous {\it inner phase shifts} $\delta_\ell(\nu,\,t)$ for any frequency $\nu$ at any acoustic radius $t_f$ in the outer layers of a star determined solely by the structure of the star interior to $t_f$, and continuous {\it outer phase shifts} $\alpha(\nu,\,\tau_f)$ that are determined solely by the structure of the outer layers above $\tau_f=T-t_f$. 

\subsection{The phase function: a main sequence example}
 
To illustrate the analysis we consider Model A, a model of a star of mass $1.15 M_\odot$ with initial hydrogen abundance $X=0.72$,  heavy element abundance $Z=0.015$, evolved to a central hydrogen abundance  $X_c=0.246$, at an age of $2.86\,10^ 9$ ys.
 The model frequency set  is shown in the echelle diagram  in Fig 1, and has 14 frequencies in the range $1482-2792\mu$Hz for each $\ell = 0,1,2,3$. The mean large separation $\Delta \sim 106.5\mu$Hz.
 
 \begin{figure}[h]
  \begin{center} 
   \includegraphics[width=8.87cm]{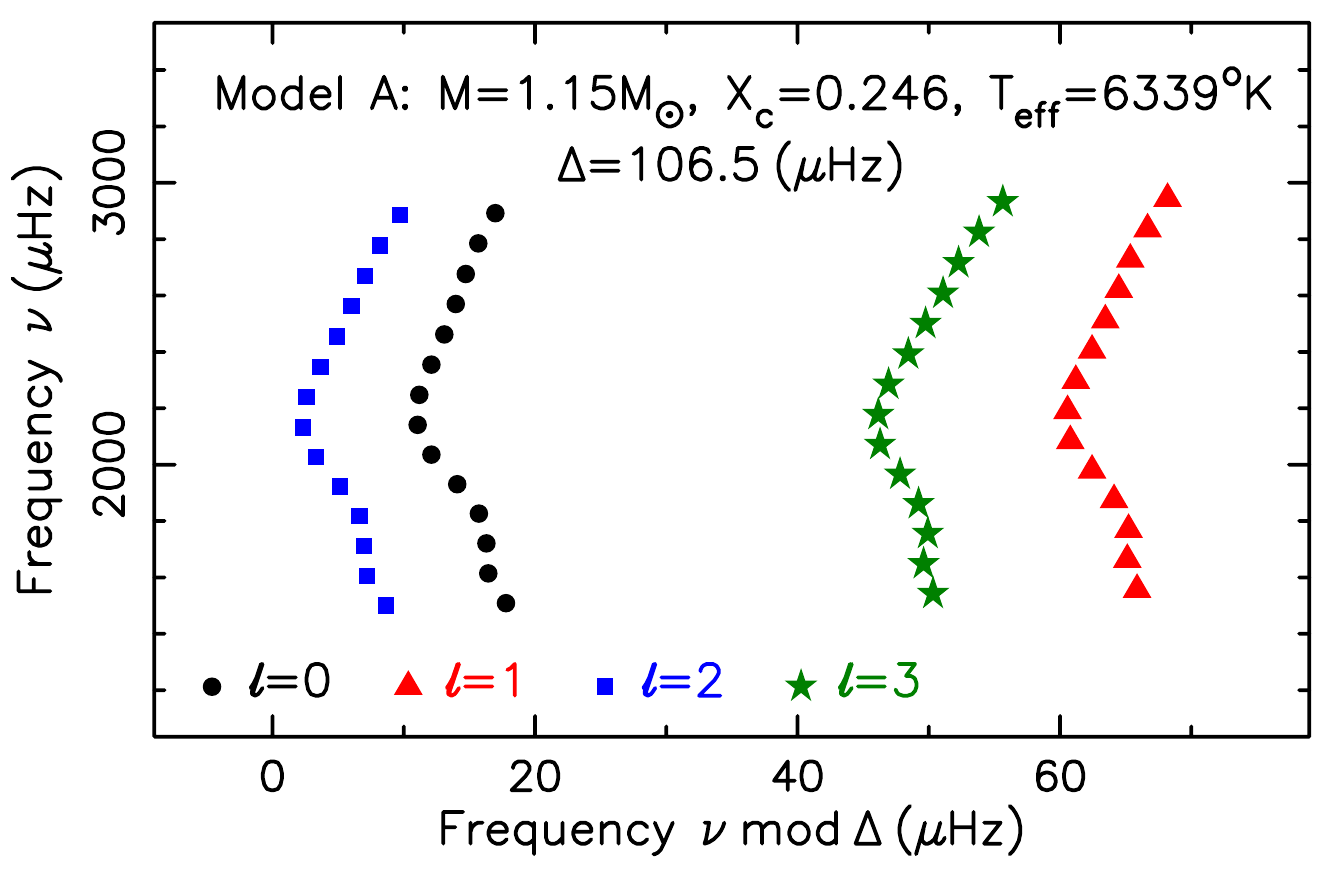}
  \end{center}  
   \vskip -14pt
   \caption{Echelle diagram of the frequencies of Model A}
\end{figure}

\begin{figure}[t]
  \begin{center} 
   \includegraphics[width=8.75cm]{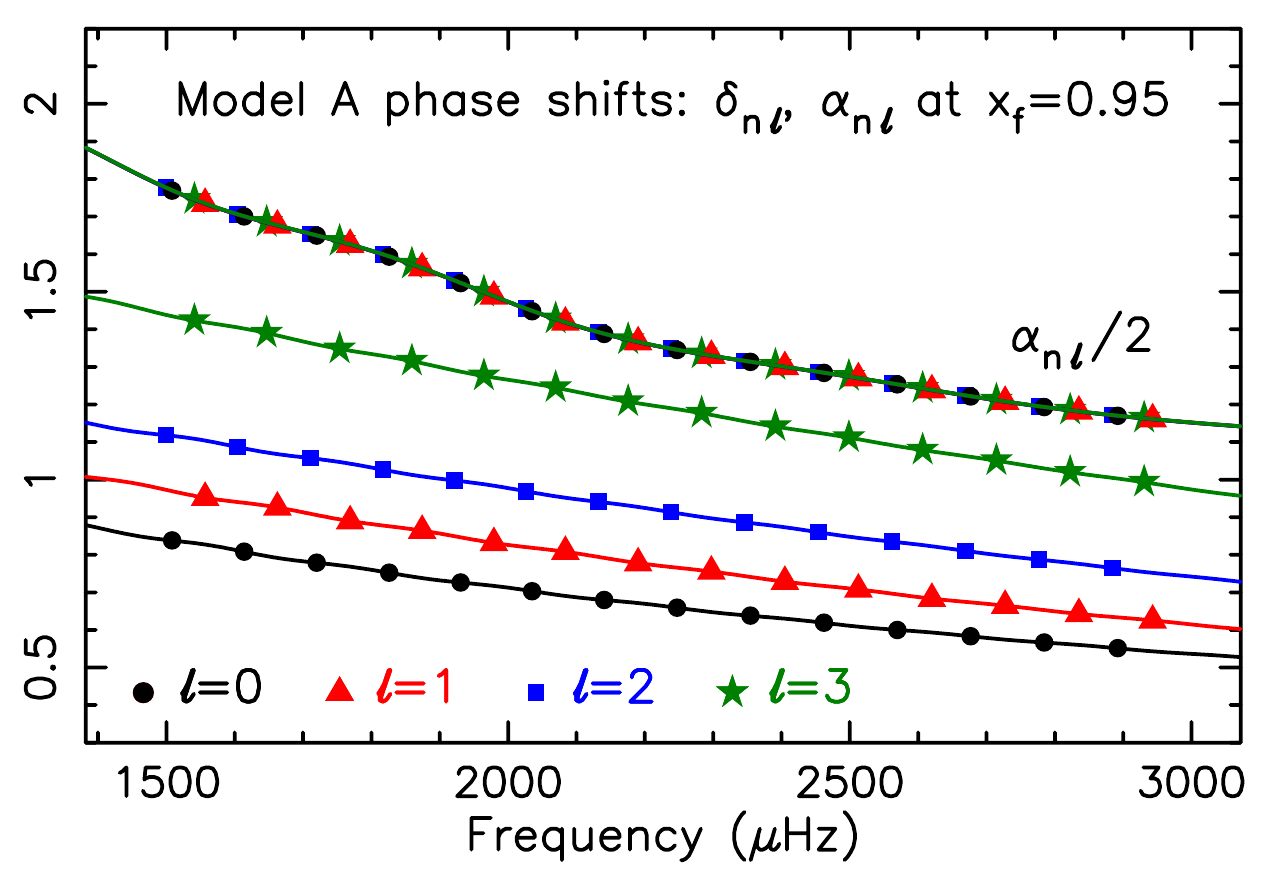}
  \end{center}  
   \vskip -16pt
   \caption{Inner and outer phase shifts  $\delta_{n\ell}, \alpha_{n\ell}$   for Model A and the continuous phase shifts 
    $\delta_\ell(\nu), \alpha_\ell(\nu)$.}
       \vskip-5pt
\end{figure}
\begin{figure}[t]
  \begin{center} 
   \includegraphics[width=8.75cm]{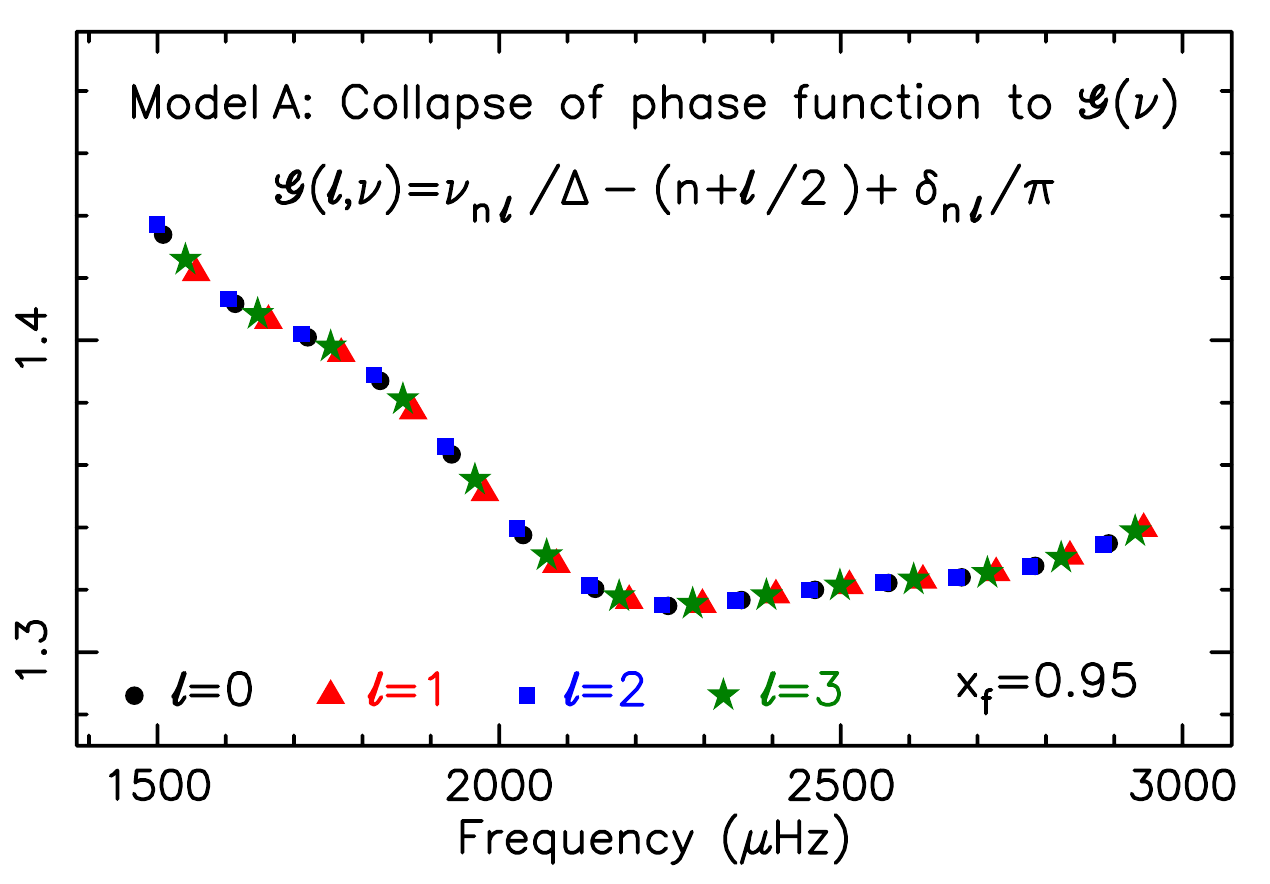}
  \end{center}  
   \vskip -16pt
   \caption{The phase function ${\cal G} (\ell, \nu)$ for Model A.}
   \vskip-10pt
\end{figure}
 
 In  Fig 2 we show these phase shifts for Model A evaluated at a fractional radius $x_f=r/R=0.95$.  
 The $\delta_{n\ell}$ depend on both degree $\ell$ and radial order $n$, whereas the $\alpha_{n\ell}$ almost all lie on the same curve $\alpha(\nu)$ independent of degree $\ell$,
  the maximum departure $\delta\alpha_\ell/\alpha(\nu)\sim4\,10^{-4}$ (see Appendix).  It is this $\ell$ independence that is the basis of the phase matching technique (and of the use of separation ratios). The curves in his figure are the continuous phase shifts calculated as described in the previous section.

The acoustic radius $T$ is not an observable, so we write Eqn 7  in terms of the observable mean large separation $\Delta$ (eg the value $106.5\mu$Hz  in Fig 1) and a {\it phase function} ${\cal G}(\ell,\nu)$ defined by
  $$ {\cal G}(\ell,\nu)\equiv
{ \nu_{n\ell}\over\Delta} - (n + \ell/2)  + {\delta_{n\ell}\over\pi} 
 = {\alpha_{n\ell}\over\pi} + \nu_{n\ell} \left( {1\over\Delta} - {2 T}\right)\eqno(8)$$
 
Taking  $\alpha_{n\ell}(\nu)$ to be an  $\ell$ independent function of $\nu$, and the last term just a linear function of $\nu$, it follows that ${\cal G}(\ell, \nu)$ is 
independent of $\ell$. This is shown in Fig 3 for our model star. With an assumed error on the frequencies of $0.2\mu$Hz the reduced $\chi^2$ for a fit to a function only of $\nu$ at $x_f=0.95$ is $0.01$, and is smaller if $x_f$ is taken closer to the surface since $\delta\alpha_\ell/\alpha(\nu)$  decreases with increasing $x_f$ (see Appendix).


This result provides the basis of model fitting by phase matching. If the model star has the same interior structure as an observed star then the function of the internal phase shifts ${\cal G}(\ell,\nu)$ of the model calculated using the {\it observed} frequencies should collapse to a single function of frequency.  The extent to which this is satisfied is a measure of the goodness of fit.

\begin{figure}[t]
  \begin{center} 
   \includegraphics[width=8.87cm]{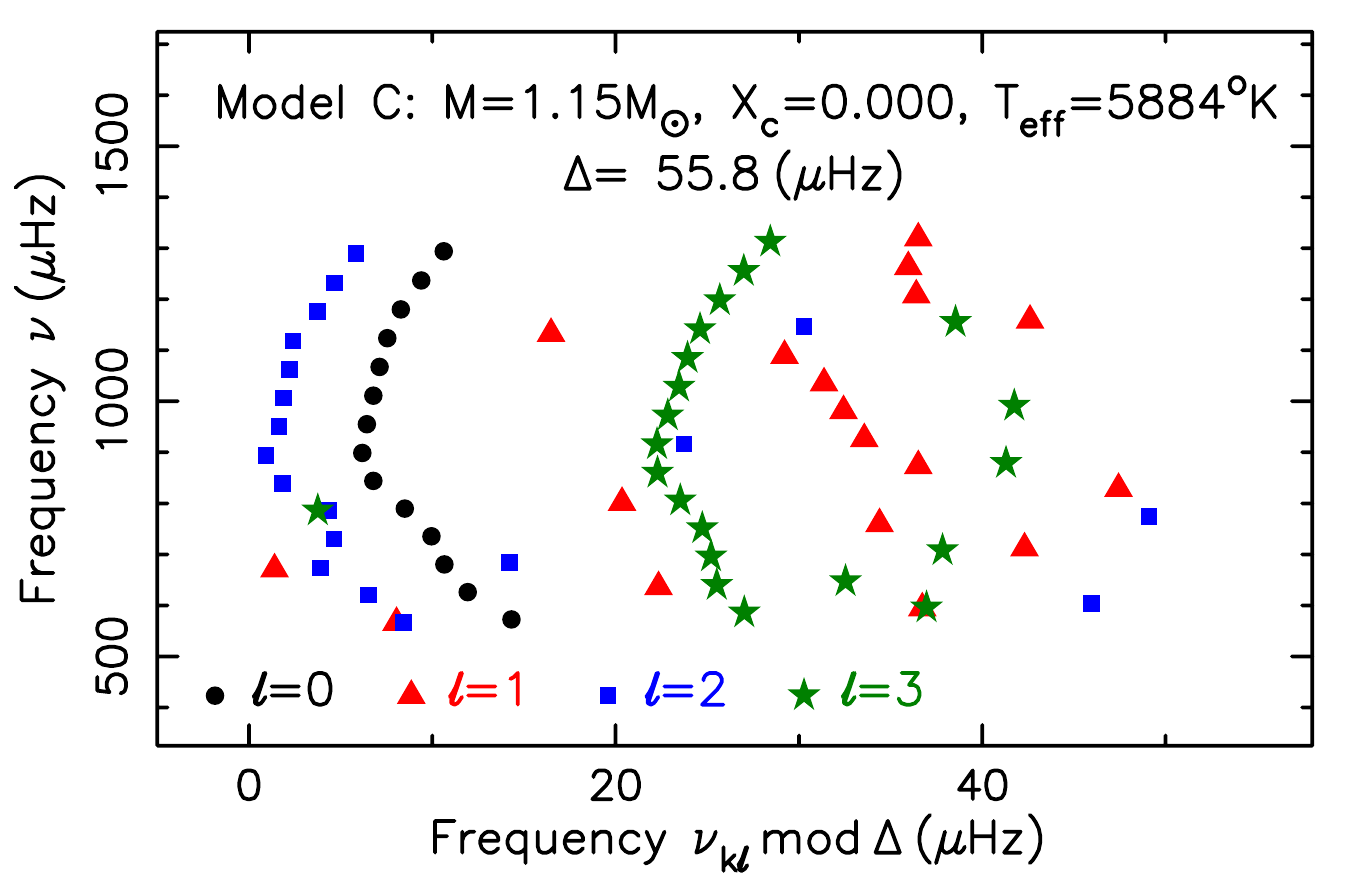}
  \end{center}  
   \vskip -16pt
   \caption{Echelle diagram of the frequencies of Model C}
  \begin{center} 
  \vskip 4pt
   \includegraphics[width=8.75cm]{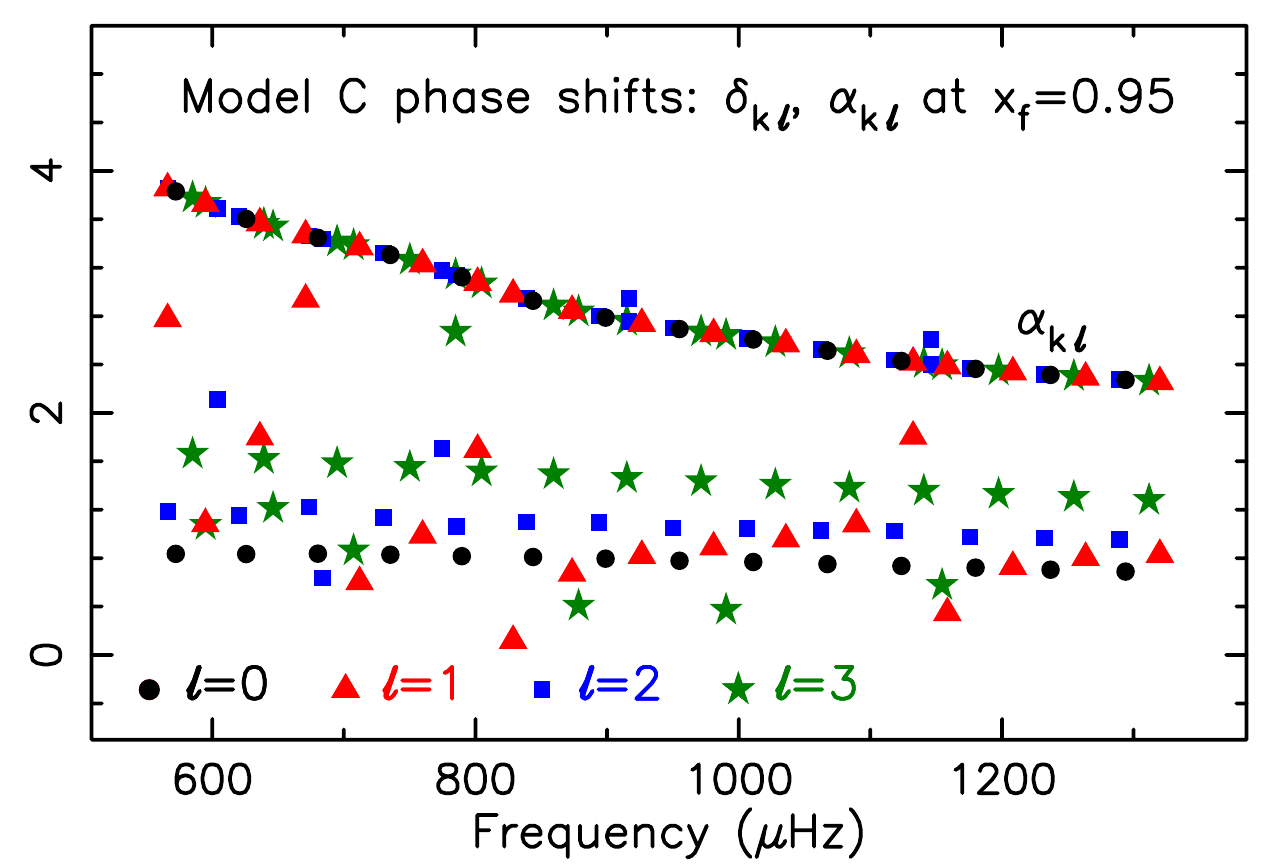}
  \end{center}  
   \vskip -16pt
   \caption{Inner and outer phase shifts  $\delta_{n\ell}, \alpha_{n\ell}$   for  Model C}
\vskip 22pt
  \begin{center} 
   \includegraphics[width=8.75cm]{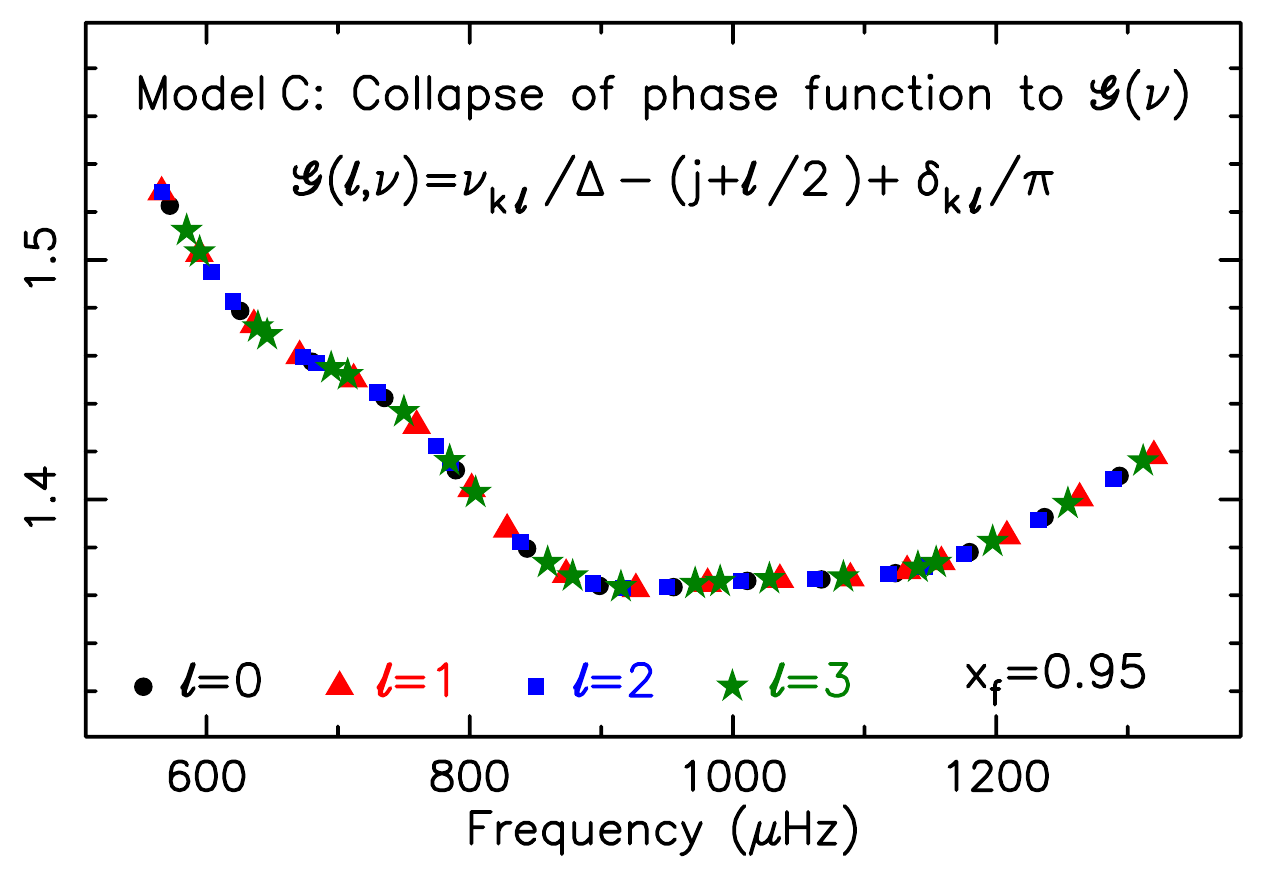}
  \end{center}  
   \vskip -14pt
   \caption{The phase function ${\cal G} (\ell, \nu)$ for Model C.}
   \vskip-18pt
\end{figure}

\subsection{Mixed modes}
The above is all for p-modes where the frequencies all follow a regular pattern and the $n$ 
value is related to the radial order of a mode - essentially the number of nodes in the eigenfunction.  
But for later stages of evolution where there are mixed modes the situation is more complicated.  
We give an example of such a star, Model C, which is  well beyond the main sequence and into the hydrogen shell burning phase as the star moves over to the red-giant branch in the H-R diagram.   The echelle diagram for frequencies in the range  $566 - 1320\mu$Hz is shown  in Fig 4, there are many mixed modes. The subscript $k$ is here just a label
 to distinguish different frequencies as $\nu_{k\ell}$.

One can of course still calculate the inner and outer phase shifts, $\delta_{k\ell}, \alpha_{k\ell}$  of the $\nu_{k\ell}$
and the they still necessarily satisfy Eqn 7, with $n$ some integer.  The phase shifts are shown in Fig 5 . As the phase shifts are are only determined to within a multiple of $\pi$ we have added or subtracted an integral number of $\pi$ to bring all the inner phase shifts within the interval $\{0,\pi$\}. The outer phase shifts $\alpha_{k\ell}$ still lie on a single curve 
$\alpha(\nu)$.

To calculate the phase function we define the integer $j$ by
 $$ j = Int \left( { \nu_{k\ell}\over\Delta} -  \ell/2  +  {\delta_{k\ell}\over\pi}           \right) \eqno(9)$$
 and the phase function by 
  $$ {\cal G}(\ell,\nu)\equiv
{ \nu_{k\ell}\over\Delta} - (j + \ell/2)           + {\delta_{k\ell}\over\pi}  \eqno(10)$$
The resulting ${\cal G}$ is shown in Fig 6. It collapses to a single function of  frequency.

  \subsection{Model fitting by phase matching}
  We here summarise the phase match algorithm.
 
Given a set of observed frequencies  $\nu^o_{n\ell}$ with error estimates $\sigma^o_{n\ell}$, and  a model to be tested for goodness of fit, 
determine the inner phase shifts $\delta_\ell(\nu^o_{n\ell})$  of the model  at the {\it observed} frequencies (Eqn 6a) and hence the phase function ${\cal G}(\ell,\nu^o_{n\ell})$ for that model (Eqn 8),  where $\Delta$ is any estimate of the average large separation of the  $\nu^o_{n\ell}$.  

Repeat this calculation for frequencies 
{ $\nu^o_{n\ell} + \sigma^o_{n\ell}$}  and subtract to determine error estimates  $e_{n\ell}$ on the determination of ${\cal G}(\ell, \nu^o_{n\ell}) $.

Then determine ${\cal A}_M(\nu)$,  a best fit  function of frequency to the  ${\cal G}(\ell, \nu^o_{n\ell})$ with errors $e_{n\ell}$  ($M$ being the number of parameters in ${\cal A}_M$), and determine the goodness of fit in terms of
 $$\chi^2 =  {1\over N-M} \sum_{n,\ell}{ \left( {\cal G }(\ell, \nu^o_{n\ell}) -  {\cal A}_M (\nu^o_{n\ell})   \over e_{n\ell}\right)^2 }\eqno(11)$$
where N is the number of frequencies.  In practice ${\cal A}_M(\nu)$ is modelled as a series in Chebyshev polynomials and $M\le N_0$ is taken to minimise
 $\chi^2$,  where $N_0$ is the number of $\ell=0$ frequencies.  

All models with $\chi^2\le1$ have an {\it interior} structure that is consistent with the observed frequencies..  Surface  layer independent model fitting algorithms necessarily cannot give information on the outer layers of the star.

To determine the $\delta_\ell(\nu)$ for any $\ell$ and $\nu$, one obtains the solution of the oscillation equations for this $\nu$ and $\ell$ which satisfy the the 
boundary condition  of regularity at $t=0$, and the surface boundary condition that the gravitational potential perturbation $\phi'_\ell$ and its derivative are continuous with a solution of Laplace's equation,  $\phi'_\ell\propto r^{-(\ell+1)}$. 
 In general this consists of calculating the two independent solutions from the centre and combining them to satisfy the Laplace condition; this can be done at any point in the outer layers since these low density regions make a negligible contribution to $\phi'$. 
One then forms the scaled Eulerian pressure perturbation 
  $\psi_\ell=r\,p'/(\rho c)^{1/2}$  and its derivative $ d\psi_\ell / dt  $ at some point $t_f$ (eg where $ x_f=0.95$).   The inner phase shifts can then be determined (for example) by

  $$\delta_{\ell}(\nu) =  \tan^{-1}\left( A - B\over 1+A B\right) \eqno(12)$$
  where 
$$A= \left(2\pi\nu\psi_\ell\over d\psi_\ell/dt\right)_{t_f },~~~
B=\tan(2\pi\nu\,t _f- \ell\pi/2) \eqno(13)$$

We emphasise that the surface phase shift $\alpha(\nu)$ of the model does not enter into this algorithm, all that is assumed is that the unknown surface phase shift  can be modelled by a single function of  frequency.

One may of course add other restrictions on the models: mass, luminosity, radius, effective temperature, surface gravity, surface composition \dots  but these are not tested by surface layer independent model fitting,  

\subsection{An example}
To demonstrate the phase matching technique we take the frequencies  of  Model A as the {\it observed} frequencies $\nu^o_{n\ell}$, with an assumed error estimate $\sigma^o_\nu= 0.2\mu$Hz on all  frequencies,
and test for a fit to other models in the same evolutionary sequence. The first example is  Model B,  the $1.15M_\odot$ model evolved to a  central hydrogen abundance of $X_c=0.009$ at an age of 
$4.52\,10^9$ys.
 
 The results are shown in Fig7 ;  ${\cal G}(\ell, \nu^o_{n\ell})$ does not collapse to a function only of $\nu$, The continuous curve is the least squares fit to a single function of $\nu$, the error estimates on ${\cal G} \approx \sigma/\Delta =1.9\,10^{-3}$  are too small to be seen in this figure. The $\chi^2$ of the fit is  $136$; the {\it observed} frequencies of model A do not fit model B. 
  
  \begin{figure}[h]
  \begin{center} 
   \includegraphics[width=8.75cm]{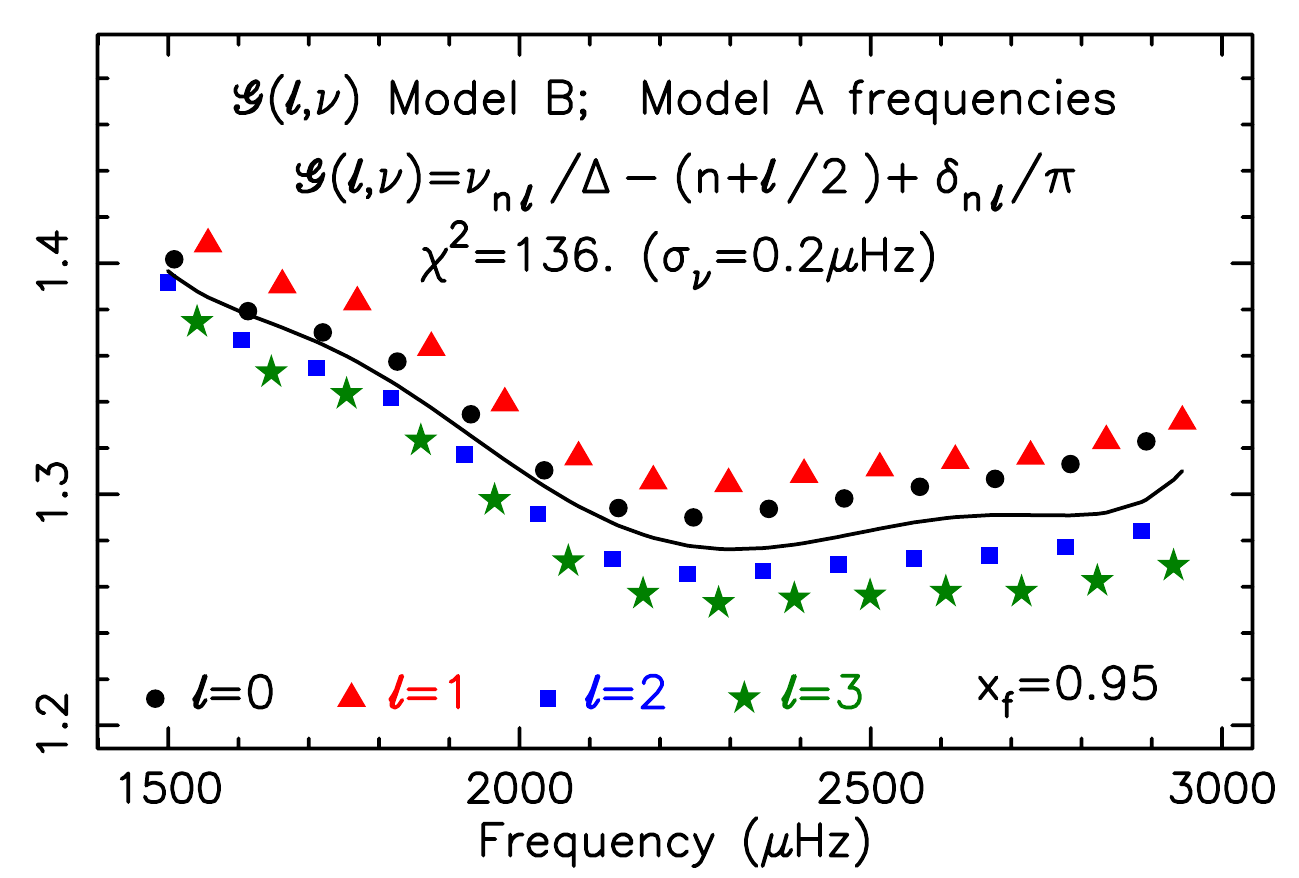}
  \end{center}  
   \vskip -16pt
   \caption{${\cal G} (\ell, \nu)$ for model B at the frequencies of model A}
   \vskip-20pt
\end{figure}

 \begin{figure}[h]
  \begin{center} 
   \includegraphics[width=8.75cm]{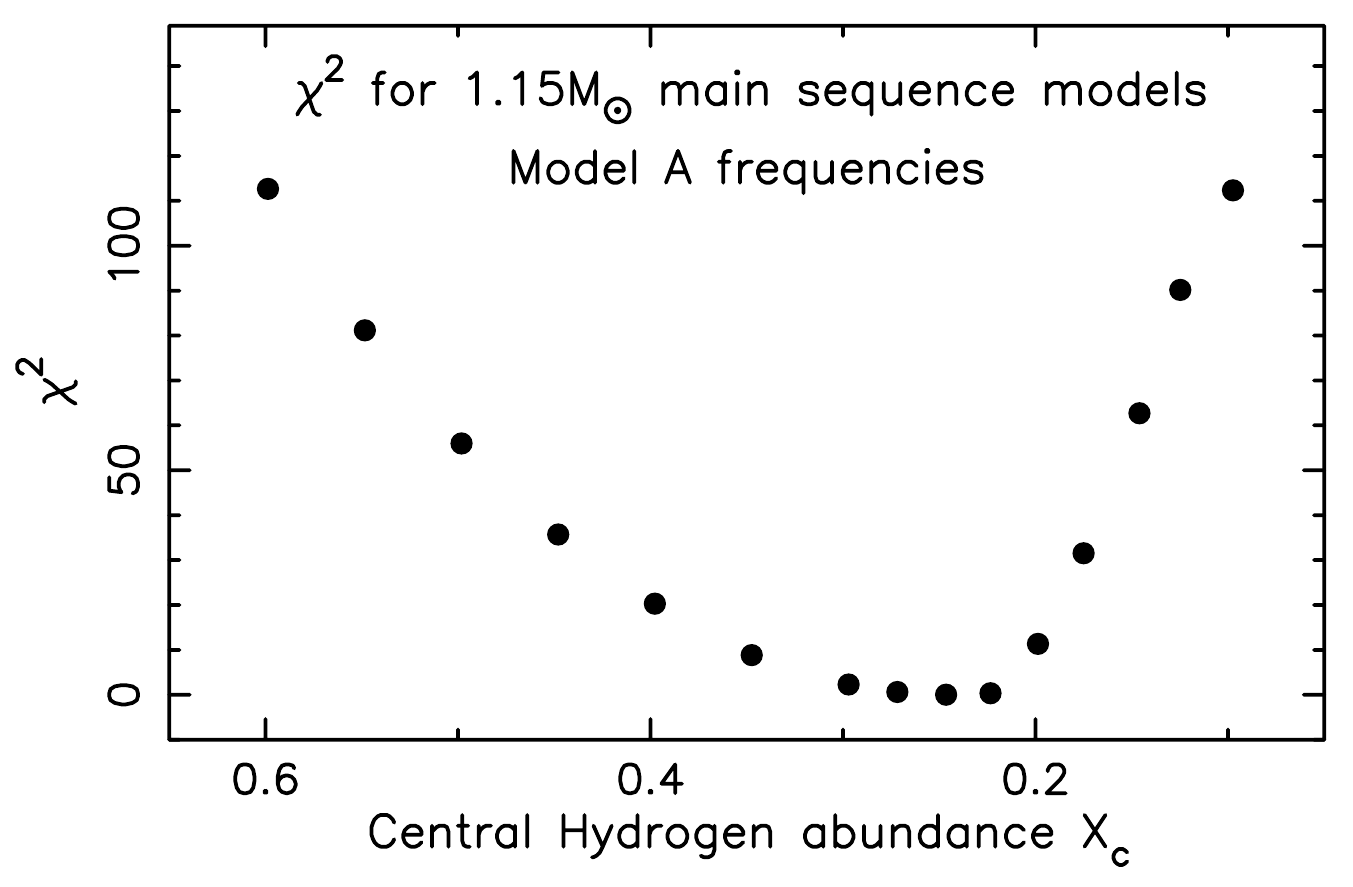}
  \end{center}  
   \vskip -12pt
   \caption{~$\chi^2$ for models in the $1.15M_\odot$ evolutionary sequence. }
   \vskip-10pt
\end{figure}
We then repeated this analysis for a set of models in the evolutionary sequence of the $1.15M_\odot$ star from the initial main sequence $X_c=0.718$, to terminal main sequence with $X_c=10^{-5}$ at an age of $5.02\,10^9$ys.  Fig 8 shows the resulting $\chi^2$ for this series of models.  Obviously models that are far removed from Model A have a very poor fit,  so to examine more closely the models in the neighbourhood of ModelA we used a finer output mesh in the evolutionary calculation and the models with 
  $\chi^2 <  1$  are shown in Fig 9. With an estimated error on the frequencies of $0.2\mu$Hz all these models fit the {\it observed} input frequencies..
 
 To further restrict the models we need to impose additional constraints. As pointed out in Roxburgh (2014), since the objective of surface layer independent model fitting is to subtract out  the effect of the outer layers it is inconsistent to impose  a constraint that the model large separation fits the observed value since the outer layers of a star make a major contribution to the large separations.  To a lesser extent the same is true of the radius, but the luminosity of the star is determined solely by the interior structure so it is reasonable to require the the model luminosity fits the observed value - in this case the luminosity of Model A. 
 
 For an increasing number of stars bolometric flux measurements and parallaxes give the luminosity to high precision, often better that $1\%$ ( cf Boyajian et al, 2013). In Fig 9 the red squares match the luminosity of ModelA to within $0.5\%$ whilst the blue triangles match to within $1 \%$ - much restricting the models that fit the input model.

  \begin{figure}[h]
  \vskip-6pt
  \begin{center} 
   \includegraphics[width=8.75cm]{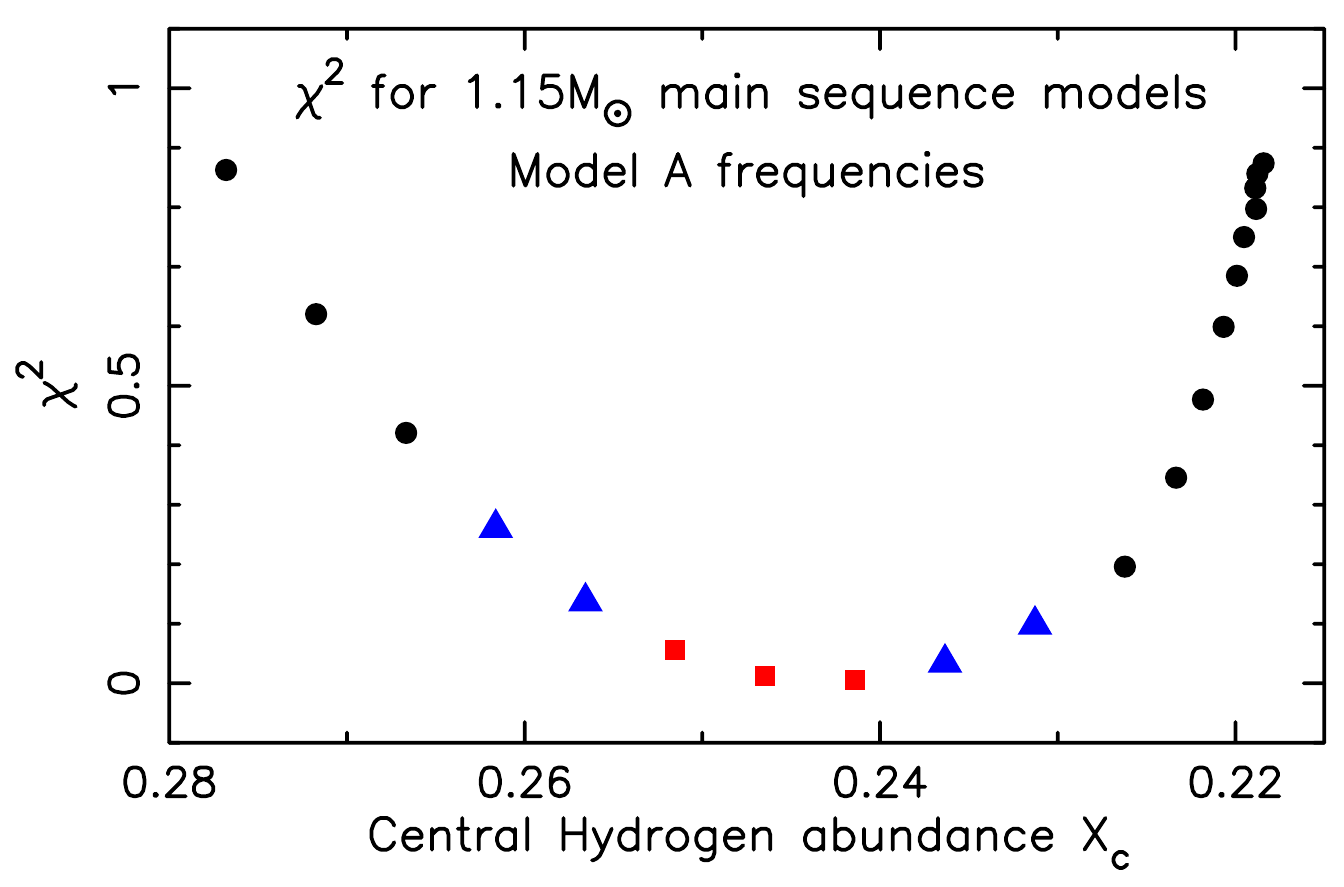}
  \end{center}  
   \vskip -14pt
   \caption{~$\chi^2$ for models in the $1.15M_\odot$ evolutionary sequence. }
   \vskip-10pt
\end{figure}


\section{Application to  HD 49933 and HD 177513}
\subsection{Constraints on model fitting}
Since the objective of phase matching model fitting (and model fitting using the ratios of small to large separations) is to subtract off the unknown effect of the outer layers of a star and find the best interior model, it would be inconsistent to constrain the model search by observed parameters that depend on the structure of the outer layers.  The two truly outer layer independent constraints are the mass $M$ and luminosity $L$.  The luminosity can be estimated using the parallax and either the magnitude or measurements of the bolometric flux.  The mass may be dynamically determined for stars in a binary system (eg $\alpha$ Cen A\&B),  or from the radius $R$  (determined either by interferometry or from $L$ and spectroscopically determined $T_{eff}$) combined with a spectroscopically determined surface gravity $\log g$. 
The  atmospheric composition is clearly a surface layer parameter.  The outer layers make a large contribution to the large separation (Roxburgh 2014)  and a smaller contribution to the radius, so estimating mass from the (approximate) scaling relation for the large separation  $\Delta\propto (M/R^3)^{1/2}$  should be treated with caution.
Ideally therefore we should constrain our model search just imposing constraints on the mass and luminosity; the radius can be used whilst recognising that it is also subject to some uncertainty due to uncertainty in the structure of the outer  layers.
Were we to find in our model fitting by phase matching, constrained solely by the estimates of $L, M$, that a best fit model also fitted the observed radius, large separation and surface composition this would be a bonus - indicating that our modelling of the outer layers is reasonable. But a model that best fits the interior may not fit the outer layer constraints because the modelling of the outer layers is incorrect.

\subsection{HD49933}
The star HD 49933 was observed by CoRoT in 2 separate runs; the initial short run was described in Appourchaux et al (2008) and the combined date from the two runs in Benomar  et al (2009) who tabulated a total of 50 frequencies in the range $1200=2600\mu$Hz including modes of degree $\ell=0,1,2$. These are the frequencies used in our analysis. There is no unique way of defining the average large separation but if one wishes to compare model and observed values we should at least use the same definition for both. We therefore define $\Delta$ as the average over 5 $\ell=0$ frequencies  centred around  the frequency closest to $\nu=1800\mu$Hz, which gives a value 
$\Delta=85.4 \pm 1.0\mu$Hz, the  error estimate allowing for the quasi periodic behaviour of the large separations of HD49933 with frequency (cf Benomar et al 2009) and the difference between the observed solar value and that of solar models. 

Taking the parallax from van Leeuwen (2007),  angular diameter from Bigot et al (2011), bolometric flux from Boyajian et al (2013)   gives 
$R/R_\odot =1.42\pm 0.04 , L/L_\odot=3.508\pm 0.090, T_{eff}=6635 \pm 90 ^o$K. But it is difficult to put any reliable constraints on the mass.
Spectroscopic and photometric measurements of  $\log g$ made by several authors are summarised in Bruntt (2009) who gives his most recent values as  
$4.28\pm 0.06$ (spectroscopic) and $4.30\pm 0.15$ (photometric).  These differ very substantially from the value of $4.0\pm 0.15$ from  Kallinger et al (2010).  If we use Bruntt's spectroscopic value and Bigot's radius the $1\sigma$ range for the mass is $1.20 \le M/M_\odot < 1.62$. 
 For comparison the $1\sigma$ mass range from using the scaling relation on $\Delta$  is $1.05  < M/M_\odot <1.25$, but the scaling relation itself is, at best, only approximate, so 
 we do not use this as a constraint but directly compare the model and observed $\Delta$s.
These parameters are listed in Table 1.

\begin{table} [h]
\setlength{\tabcolsep}{3pt}
\centerline   {\bf Table 1.  HD 49933:  Possible constraints on model fitting}
\vskip 0.15cm
\tiny
\centering
 \begin{tabular}{c c c c c c c c c c c c c c c   } 
\hline\hline 
\noalign{\smallskip}
\tiny
 $L/L_\odot$&$M/M_\odot$&$R/R_\odot$  &$\Delta$    \\	[1ex]
\hline 
\noalign{\smallskip}
\tiny
$3.508\pm 0.090$ & $1.41\pm0.21$&$1.420\pm 0.040$  &$85.4\pm 1.0$\\[0.2ex]
\hline
\end{tabular}
\vskip -10pt
 \end{table}
\normalsize

We searched for a best fit  model using only the phase match criterion plus the luminosity constraint. The model set was constructed using the STAROX code (Roxburgh 2008b) which
has GN93 relative abundances (Grevesse and Noels 1993), OPAL opacities (Iglesias and Rogers 1996) supplemented by Whichita opacities at low temperatures (Ferguson et al 2005), and NACRE reaction rates (Angulo et al 1999) (taking $^{13}$C,  $^{15}$N and $^{17}$F to be in equilibrium), the mixing length model of convection (with mixing 
length=$\alpha_MH_p$), two models of convective overshoot and mixing from a convective core, one set (IR1)  with just chemical mixing a distance of $\alpha_oH_p$, the second (IR2) with full entropy and chemical mixing a distance $\alpha_{o}d_I$ where  $ d_I$ is the distance given by the Integral Constraint (Roxburgh 1978, 1989).  These models do not include diffusion so we also search a model set (AM) kindly supplied by A Miglio (2012) computed using the Li\`ege CL\'ES code (Scuflaire et al 2008) with and without chemical overshooting and diffusion. 
The model set IR1 has 100,440 models with mass $M/M_\odot =0.9-1.50$, initial composition $X_H=0.68-0.74$, $Z=0.009- 0.024$,  $\alpha_M=1.6-2.4 $, and $\alpha_o=0-0.2$. Model set IR2 has  $33,480$ models with the same parameter set but $\alpha_o=0.5$

A very large number of models were found to fit the both the phase match and luminosity constraints to within a $\chi^2 \le 1$; $446$ from set IR1, and $138$ from IR2, with masses ranging from $1.10-1.44 M_\odot$ and ages from $1.12 -3.97\,10^9$ys, 
all with some core overshooting. Imposing the mass constraint reduces this slightly  to $379$ and $123$, restricting the mass range to $1.20\le M/M_\odot\le 1.44$.
Imposing the radius constrain reduces this to  $123$ for IR1 and $40$ for IR2 but still with a similar range in mass and age. Finally imposing the large separation constraint reduces the models to 7 from IR1, 2 from IR2 and 4 from AM:  details  of these models are given in Table 2. Figure 10 shows the phase match function ${\cal G}(\ell,\nu)$ for one of  the IR2 models. No models with diffusion and/or no overshooting satisfied all the constraints. 

\begin{table} [h]
\setlength{\tabcolsep}{1.5pt}
\centerline   {\bf Table 2:  Best fit models HD49933:  IR1, IR2, AM model sets.}
\vskip 0.15cm
\tiny
\centering
 \begin{tabular}{l c c c c c c c c c c c c c c   } 
\hline\hline 
\noalign{\smallskip}
\tiny
Set&$M/M_\odot$ & $~L/L_\odot$ & $~R/R_\odot$ & $~~XH~~~$ &$~~~~Z~~~$& $~~~\alpha_M~$ & $\alpha_o~~$ & $~~age~~$& $~~~~\Delta~~$ & $~~~~X_c~~$ &$~~~~\chi^2_\nu$    \\	[1ex]
\hline 
\noalign{\smallskip}
\tiny
IR1 &1.26&$3.58$ & $1.46$ & $0.72$ & $0.015$ & $2.0$ &  $0.2$ &    $2.76$ & 86.2 & 0.39 &0.61\\[0.2ex]
IR1 &1.22&$3.45$ & $1.46$ & $0.68$ & $0.018$ & $2.0$ &  $0.2$ &    $2.77$ & 85.5& 0.35 &0.65\\[0.2ex]
IR1 &1.20&$3.46$ & $1.44$ & $0.72$ & $0.012$ & $2.0$ &  $0.2$ &    $3.21$ & 86.3 & 0.35 &0.69\\[0.2ex]
IR1 &1.22&$3.56$ & $1.46$ & $0.68$ & $0.018$ & $2.4$ &  $0.2$ &    $3.06$ & 85.3& 0.30 &0.80\\[0.2ex]
IR1 &1.20&$3.57$ & $1.45$ & $0.68$ & $0.015$ & $2.0$ &  $0.1$ &    $2.63$ & 84.7 & 0.29 &0.84\\[0.2ex]
IR1 &1.22&$3.58$ & $1.45$ & $0.68$ & $0.015$ & $1.6$ &  $0.1$ &    $1.98$ & 84.4.& 0.39 &0.86\\[0.2ex]
IR1 &1.22&$3.48$ & $1.46$ & $0.70$ & $0.015$ & $2.0$ &  $0.1$ &    $2.85$ & 85.4& 0.29 &0.88\\[0.2ex]
IR2 &1.24&$3.59$ & $1.46$ & $0.68$ & $0.018$ & $2.0$ &  $0.5$ &    $2.51$ & 85.7 & 0.39 &0.67\\[0.2ex]
IR2 &1.22&$3.51$ & $1.46$ & $0.72$ & $0.012$ & $2.0$ &  $0.5$ &    $2.51$ & 85.7 & 0.35 &0.73\\[0.2ex]
AM &1.22&$3.49$ & $1.44$ & $0.69$ & $0.015$ & $1.9$ &  $0.2$ &    $2.71$ & 86.2 & 0.38 &0.61\\[0.2ex]
AM &1.26&$3.45$ & $1.45$ & $0.71$ & $0.015$ & $1.7$ &  $0.2$ &    $2.40$ & 86.3 & 0.45 &0.66\\[0.2ex]
AM &1.24&$3.57$ & $1.45$ & $0.69$ & $0.015$ & $1.7$ &  $0.2$ &    $2.24$ & 85.2 & 0.43 &0.66\\[0.2ex]
AM &1.22&$3.51$ & $1.43$ & $0.64$ & $0.015$ & $1.7$ &  $0.2$ &    $2.23$ & 85.7 & 0.42 &0.66\\[0.2ex]
\hline\hline 
\noalign{\smallskip}
IR1 &1.10&$3.47$ & $1.39$ & $0.68$ & $0.009$ & $2.0$ &  $0.2$ &    $3.12$ & 86.3 & 0.29 &0.80\\[0.2ex]
AM &1.30&$3.50$ & $1.46$ & $0.68$ & $0.020$ & $1.7$ &  $0.2$ &    $1.44$ & 86.6 & 0.52 &0.81\\[0.2ex] 
\hline
\end{tabular}
\vskip -10pt
 \end{table}
\normalsize

  \begin{figure}[h]
  \vskip-6pt
  \begin{center} 
   \includegraphics[width=8.75cm]{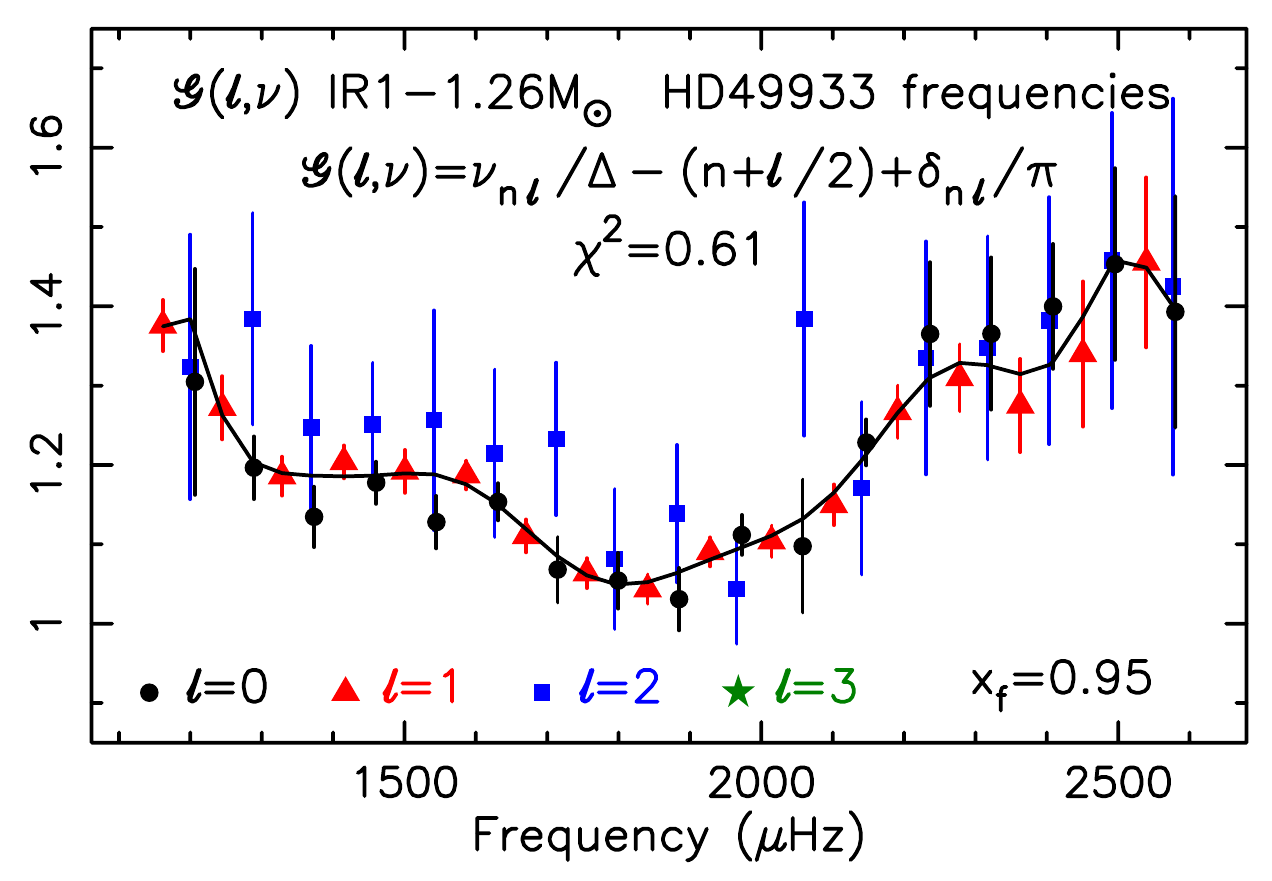}
  \end{center}  
   \vskip -14pt
   \caption{The fit of $\cal{G}(\ell, \nu)$  to a function only of frequency, $1.26M_\odot$ model to HD49933 frequencies.}
   \vskip-10pt
\end{figure}

However if we drop the constraint on the mass from $\log\,g$  and relax the constraint on $\Delta$ just a little then we obtain many more models; two examples are given at the end  of Table 2 - the AM model includes diffusion. 

The obvious conclusion from these results is that the precision on the frequencies is not good enough to determine a best fit model.  The wide range of models that fit the data are compatible with the values found by  Piau et al (2009), Benomar et al (2010), Kallinger et al (2010), Bigot et al (2011), Creevey and Bazot (2011) and Liu et al (2014).

\subsection{ HD 177153 (aka Perky)}
The star HD 177153 was observed by Kepler (KIC 6106415, aka Perky) and analysed by Silva Aguire et al (2013) who list a total of 33 frequencies in their Table 1, 11 for each 
$\ell=0,1, 2$; these are the frequencies used in this analysis. Silva Aguire et al  searched for best fit models by two procedures,: comparing frequencies with a "surface offset" 
(cf Kjeldsen et al 2008),  and  comparing separation ratios (cf Roxburgh and Vorontsov 2003, 2013) - in fact a subset of ratios (but see comments below). Here we seek best fit models by phase matching. and compare our results with theirs. 

These authors also list global parameters which we reproduce in Table 3 and use in our analysis, determining the radius constraint from $L, T_{eff}$ and the mass constraint from the radius and $\log\,g$.  We determine the large separation from the 5 $\ell=0$ frequencies centred on $2221\mu$Hz, not from the autocorrelation of the time series  (cf Roxburgh and Vorontsov 2006) as one cannot compute model values by this technique,  and we take an enhanced error estimate on $\Delta$ as was done for HD49933. 

\begin{table} [h]
\setlength{\tabcolsep}{3pt}
\centerline   {\bf Table 3.  HD 177513  Possible constraints on model fitting}
\vskip 0.15cm
\tiny
\centering
 \begin{tabular}{c c c c c c c c c c c c c c c   } 
\hline\hline 
\noalign{\smallskip}
\tiny
 $\log(L/L_\odot)$&$T_{eff}$   &$\log\,g$& $R/R_\odot$&$M/M_\odot$  &$\Delta$    \\	[1ex]
\hline 
\noalign{\smallskip}
\tiny
$0.26\pm 0.04$ & $6000\pm200$&$4.27\pm 0.10$&$1.25\pm0.10$&$1.09\pm0.3$  &$104.25\pm 1.0$\\[0.2ex]
\hline
\end{tabular}
\vskip -8pt
 \end{table}
\normalsize

We then searched  our model sets to find models that fitted the above constraints, using the full set of 33 frequencies,  and for which the phase function $\cal{G}(\ell,\nu)$ collapsed to a function only of $\nu$ within a $\chi^2 < 1$ (Eqn 9).   We included  the model set IR0 which is the same as IR1 except that it follows the evolution of $^{13}$C,  $^{15}$N and $^{17}$F.  

Again there are many models (77) that fit the constraints without fitting the large separation, whose value is  sensitive to the detailed structure of the outer layers so should not logically be imposed as a constraint on surface layer independent model fitting. If we nevertheless impose this constraint along with that of the luminosity, radius (or effective temperature) and 
mass (from $\log\,g$) we find 20 models that satisfy these constraints and also the phase match condition; these are listed in Table 4. All best fit models listed  have no overshooting: models IR0 have no diffusion; models AM all have diffusion (see Scuflaire 2008). The fit for one such model  (IR01 $1.09M_\odot$ ) is shown in Fig 11.

The models cover a mass range of $1.08$ to $1.15M_\odot$ with ages in the range $4.13$ to $5. 09\,10^9$ys.  
This is very similar to the range of models in Tables $5$ and $6$ in Silva Aguire et al (2013).  which has a mass range $1.09-1.17M_\odot$ and ages varying between $3.8$ to $5.5\,10^9$ys, although only 2 of their models have a $\chi^2 < 1$.

\begin{figure}[h]
  \vskip-6pt
  \begin{center} 
   \includegraphics[width=8.75cm]{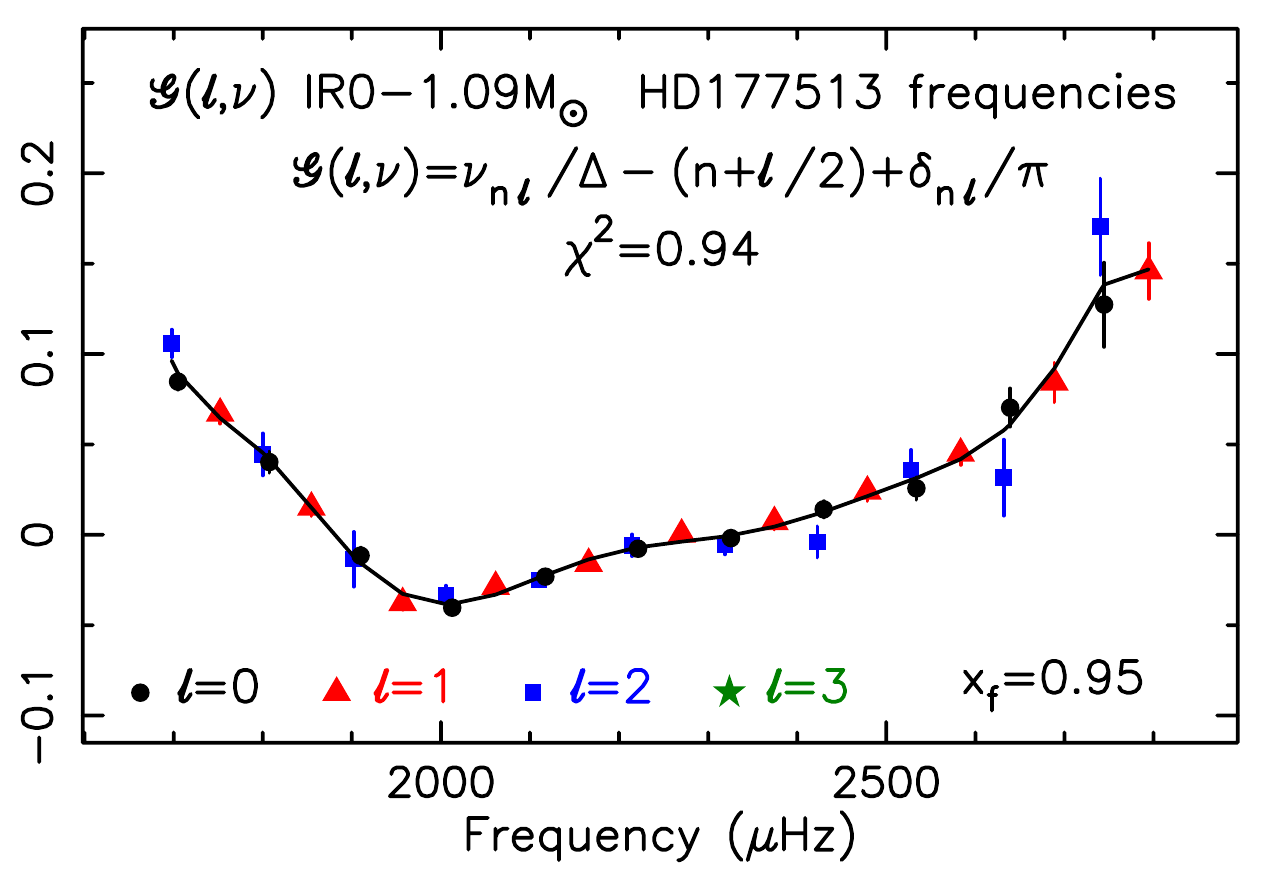}
  \end{center}  
   \vskip -14pt
   \caption{The fit of $\cal{G}(\ell,\nu)$  to a function only of frequency, $1.09M_\odot$ model to HD177513 frequencies.}
\end{figure}

\begin{table} [h]
\setlength{\tabcolsep}{1.5pt}
\centerline   {\bf Table 4:  Best fit models HD177513:  IR0, AM model sets.}
\vskip 0.15cm
\tiny
\centering
 \begin{tabular}{l c c c c c c c c c c c c c c   } 
\hline\hline 
\noalign{\smallskip}
\tiny
Set&$M/M_\odot$ & $~L/L_\odot$ & $~R/R_\odot$ & $~~XH~~~$ &$~~~~Z~~~$& $~~~\alpha_M~$  & $~~age~~$& $~~~~\Delta~~$ & $~~~~X_c~~$ &$~~~~\chi^2_\nu$    \\	[1ex]
\hline 
\noalign{\smallskip}
\tiny
IR0&1.08&  1.70&  1.23&  0.70&  0.020&   1.6&   4.86&   104.8&   0.10&   1.00\\[0.2ex]
IR0&1.09&  1.70&  1.24&  0.70&  0.021&   1.6&   4.80&   104.5&   0.10&   0.94\\[0.2ex]
IR0&1.11&  1.70&  1.25&  0.70&  0.023&   1.6&   4.72&   103.6&   0.10&   0.94\\[0.2ex]
IR0& 1.13&  1.91&  1.24&  0.72&  0.019&   1.8&   4.48&   105.2&   0.10&   1.00\\[0.2ex]
IR0&  1.11&  1.69&  1.24&  0.72&  0.020&   1.6&   4.86&   105.2&   0.15&   0.96\\[0.2ex]
IR0&  1.14&  1.91&  1.25&  0.72&  0.020&   1.8&   4.46 &  104.9&   0.10&   0.94\\[0.2ex]
IR0&  1.12&  1.69&  1.24&  0.72&  0.021&   1.6&   4.83&   104.8&   0.14&   0.99\\[0.2ex]
IR0&  1.15&  1.90&  1.25&  0.72&  0.021&   1.8&   4.42&   104.8&   0.10&   0.93\\[0.2ex]
IR0&  1.15&  1.85&  1.25&  0.72&  0.022&   1.8&   4.61&   105.1&   0.10&   0.91\\[0.2ex]
IR0&  1.13&  1.91&  1.24&  0.73&  0.018&   1.8&   4.72&   105.2 &  0.10&   0.96\\[0.2ex]
IR0&  1.14&  1.90&  1.25&  0.73&  0.019&   1.8 &  4.70&   104.9&   0.10&   0.90\\[0.2ex]
IR0&  1.12&  1.68&  1.24&  0.73&  0.020&   1.6&   5.09&   104.9&   0.15&   0.92\\[0.2ex]
IR0&  1.15&  1.90&  1.26&  0.73&  0.020&   1.8&  4.70&   104.5&   0.10&   0.91\\[0.2ex]
IR0&  1.13&  1.68&  1.25&  0.73&  0.021&   1.6&   5.08&   104.5&   0.15&   0.98\\[0.2ex]
AM& 1.11&  1.75&  1.24&  0.71&  0.020&   1.7&   4.75&   104.7&   0.13&   0.85\\[0.2ex]
AM&  1.10&  1.66&  1.23&  0.68&  0.025&   1.7&   4.59&   105.0&   0.17&   0.88\\[0.2ex]
AM&  1.13&  1.86&  1.25&  0.68&  0.025&   1.9&   4.21&   104.8 &  0.20&   0.88\\[0.2ex]
AM&  1.12&  1.70&  1.25&  0.69&  0.025&   1.7&   4.62 &  103.8 &  0.17&   0.87\\[0.2ex]
AM&  1.15&  1.89&  1.25&  0.69&  0.025&   1.9&   4.13&   104.7&   0.22&   0.92\\[0.2ex]
AM&  1.14&  1.73&  1.26&  0.70&  0.025&   1.7&  4.53 &  103.8 &  0.18&   0.97\\[0,2ex]
\hline
\end{tabular}
\vskip -10pt
 \end{table}
\normalsize

However a  difference is that we find many different models that satisfy the constraints, all produced by the same stellar evolution code with the same physical assumptions, 
in contrast to the range of models in  Silva Aguire  (2013) which were computed using different stellar evolution codes.  There is a lesson in this: if one were to search for a best fit model using some minimisation algorithm, the multi-dimensional surface in which one is seeking a minimum can have many local minima consistent with the data and its error estimates, that is with a $\chi^2 < 1$.   It is not statistically sound to search for the minimum with the lowest value of $\chi^2<1$ and consider this to be the "correct" model; all are consistent with the data.  To discriminate between the models one needs data of greater precision. 

In their analysis Silva Aguire et al (2013) combined $N_0$ ratios $r_{01}$ and $N_1$ ratios $r_{10}$  into a single series and sought  models that fit the combined $N_0+N_1$ data points.  But the $r_{01}, r_{10}$ are not independent, one set can in principle be obtained from the other by interpolation. Half  the frequencies  are
 used in eliminating the surface phase shift $\alpha(\nu)$ so there are only $\sim N_0$ independent data (see next section), taking $N_0+N_1$ over estimates the $\chi^2$.


\section{Relation to separation ratios}
Surface layer independent model fitting, phase matching and separation ratios, are both based on the result that in the outer layers of a star the surface phase shift $\alpha_\ell(\nu)$  is, to a very good approximation, independent of $\ell$ so that $\alpha_\ell(\nu)=\alpha(\nu)$ which can be considered a continuous function of $\nu$ (see Appendix).  

For any estimate of the average large separation
 $\Delta$, the oscillation frequencies $\nu_{n\ell}$ can always be  expressed in terms of functions $\epsilon_\ell(\nu)$ in the form
$$  \nu_{n\ell} = \Delta  \left[    n + \ell/2 +  \epsilon_\ell(\nu_{n\ell})   \right]  \eqno(14)$$
where, for any set of frequencies. the functions $\epsilon_\ell(\nu)$ are known at a discrete set of values $\nu_{n\ell}$
$$\epsilon_\ell(\nu_{n\ell})= {\nu_{n\ell}\over\Delta}-n-\ell/2  \eqno(15)$$

 From the {\it Eigenfrequency Equation} (Eqns 7,8), $\epsilon_\ell(\nu)$ can be expressed in terms of the inner and outer phase shifts as
$$\epsilon_\ell(\nu)= {1\over\pi} \left[ \alpha_\ell(\nu)  + \pi \nu \left({1\over\Delta}-2T\right) \right] - { \delta_\ell(\nu)\over\pi}  \eqno(16)$$
With $\alpha_\ell(\nu)=\alpha(\nu)$ independent of $\ell$, the  term in square brackets in this equation (which we call $\alpha^*$) is a function only of $\nu$.

 The phase match algorithm uses this equation in the form 
 $${\cal{G}}(\ell,\nu)=\epsilon_\ell(\nu_{n\ell})  +{ \delta_\ell(\nu_{n\ell}) \over \pi} = {\alpha^*(\nu) \over \pi} \eqno(17)$$
 is a function only of $\nu$; the separation ratios algorithm  
 uses it to subtract off the $\alpha^*(\nu)$ by interpolating in the discrete set of the $\epsilon_\ell(\nu_{n\ell})$  for values at the same frequencies, and eliminating $\alpha^*$ by subtraction to determine the inner phase shift differences, eg
 $$\epsilon_0(\nu_{n0}) - \epsilon_\ell(\nu_{n0}) = {1\over\pi} \big[ \delta_\ell(\nu_{n0}) - \delta_0(\nu_{n0}) \big]  \eqno(18)$$
which only depend on the inner structure of the star. If one uses linear interpolation one readily obtain the result that
$$\epsilon_0(\nu_{n,0}) - \epsilon_1(\nu_{n,0})=  {   \nu_{n,0} - (  \nu_{n,1} + \nu_{n-1,1})/2  \over   \nu_{n+1,1} - \nu_{n-1,1} } =r_{01} \eqno(19) $$
$$\epsilon_0(\nu_{n,0}) - \epsilon_2(\nu_{n,0})=  {  \nu_{n,0} -\nu_{n-1,2} \over  \nu_{n,2} - \nu_{n-1.2} } =r_{02}  \eqno(20)$$
where $r_{02}, r_{01}$ (and its relative $r_{10}$) are the separation ratios.  
\begin{figure}[h]
  \vskip-6pt
  \begin{center} 
   \includegraphics[width=8.75cm]{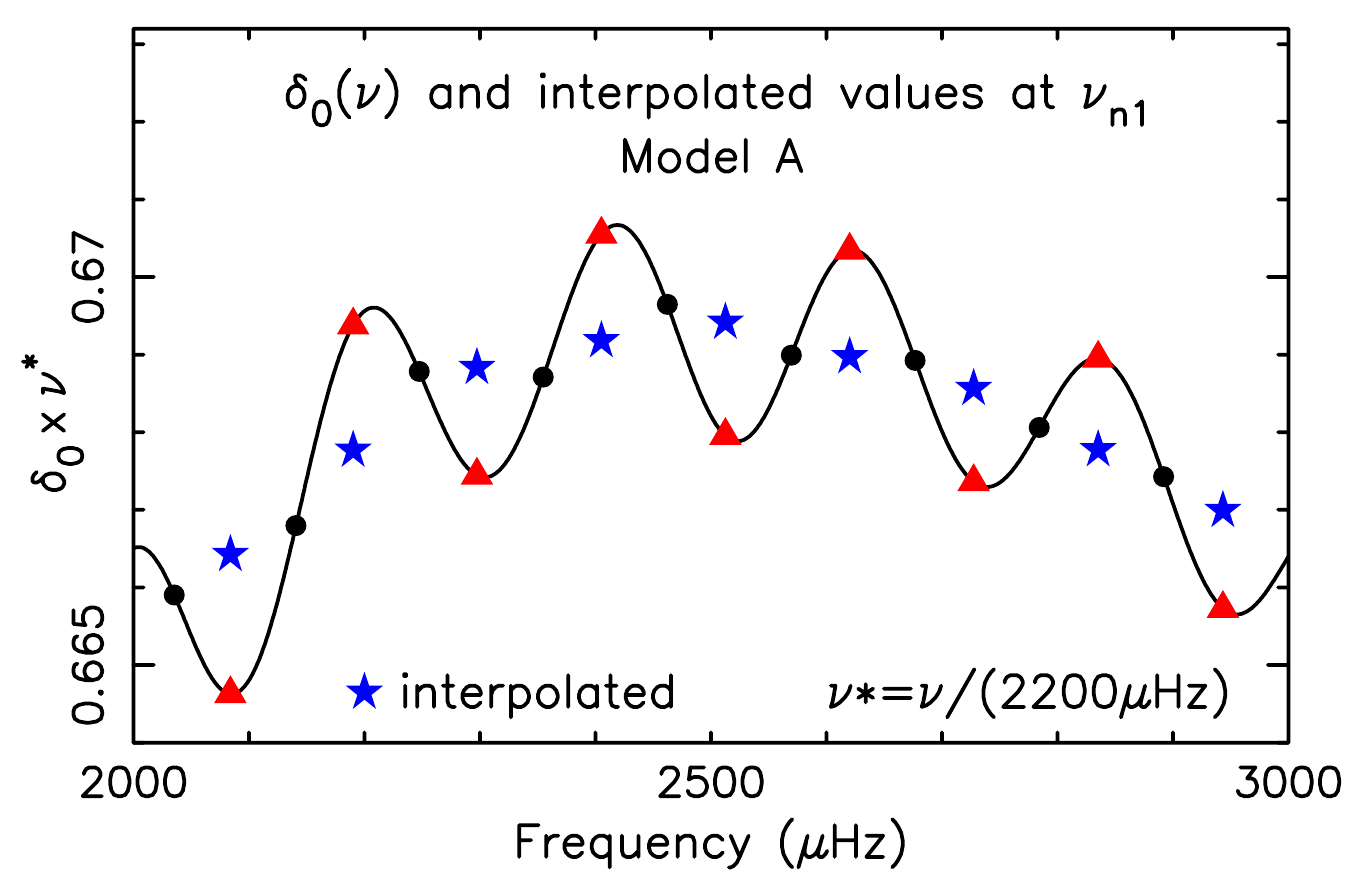}
  \end{center}  
   \vskip -14pt
   \caption{Error in determining the phase shifts $\delta_0(\nu)$ at $\nu_{n1}$ (blue stars) from the values at frequencies $\nu_{n0}$ (black points).}
   \vskip-10pt
\end{figure}

A source of error in the separation ratio technique is the error in interpolation; this can be reduced by using higher order interpolation algorithms, such as the 5 point expressions for  $r_{01}, r_{10}$ , but they still have an error which can be significant  
if  the quasi periodic modulation in the $\delta$s is on a scale smaller than twice the large separation. 
 In Fig 12 we illustrate this by interpolating for the values of $\delta_0(\nu)$ at frequencies $\nu_{n1}$ from the values at $\nu_{n0}$ for Model A. The  curve is the continuous  phase shift 
 $\delta_0(\nu)$, the black points are the discrete phase shift $\delta_{n0}$ at the frequencies $\nu_{n0}$, the red triangles the values $\delta_0(\nu)$ at frequencies $\nu_{n1}$, 
 and the blue stars the values at $\nu_{n1}$ determined from the $\delta_{n0}$ by cubic interpolation.

 \begin{figure}[t]
  \begin{center} 
   \includegraphics[width=8.75cm, height=6.cm]{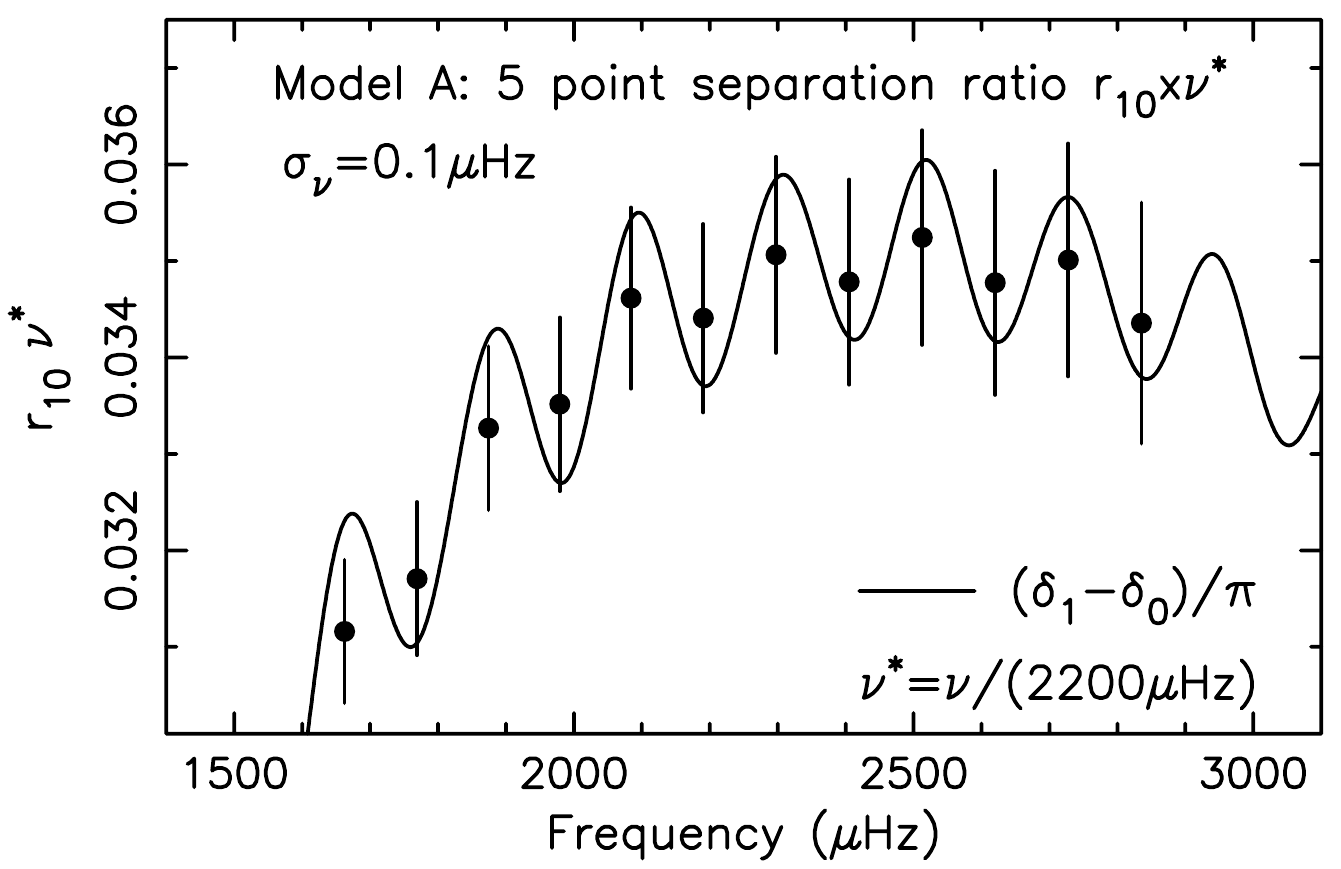}
  \end{center}  
   \vskip -14pt
   \caption{The separation ratios $r_{10}$ for Model A compared to the exact phase shift difference of [$\delta_1(\nu)-\delta_0(\nu)]/\pi$.}
   \vskip-6pt
\end{figure}

The quasi periodicity in the $\delta$s is due to the rapid variation in acoustic variable at the base of a convective envelope, the period being determined by the acoustic radius of the interface and the amplitude by the strength of the quasi discontinuity (Roxburgh (2009).  It is only an important source of error if the error from interpolation is larger than that due to errors on the frequencies.  This is not the case for Model A, Fig 13 shows the values of the 5 point small separations $r_{10}$  compared with to actual phase shift difference, the error bars corresponding to an error of $0.1\mu$Hz on the frequencies. 

We note here that one can interpolate for the value of $\epsilon_0$ and $\epsilon_\ell$ at any $\nu$ and, on subtraction, obtain  an approximation to the the phase shift difference $\delta_\ell-\delta_0$ at any $\nu$, but this does not increase the number of independent variables. If one has $N$ values of each of $\nu_{n0}$ and $\nu_{n\ell}$, then essentially $N$ of them are used to model $\alpha^*(\nu)$ leaving $N$ independent data on $\delta_0-\delta_\ell$. This is why it is incorrect to add together the series of ratios $r_{01}$ and $r_{10}$  and consider this as $2N$ data.

The phase matching technique requires that the unknown outer phase shifts $\alpha^*(\nu)$ 
can be well approximated by a polynomial in $\nu$ determined by the all the  inner phase shifts of the total number of observed frequencies. The outer phase shifts also have a quasi-periodicity determined primarily by the acoustic depth of the HeII ionisation zone, but this period is typically many large separations and so can be accurately modelled by a polynomial in $\nu$. A further advantage of the phase match technique is that  in the fitting process the errors are uncorrelated, whereas in the separation ratios technique the errors on the ratios are strongly correlated.

\section*{Appendix\\ Accuracy of the approximation {\boldmath$\alpha_\ell(\nu)=\alpha(\nu)$}} 
Surface layer independent model fitting techniques, phase matching or separation ratios, are based on the approximation that the frequency dependent outer phase shifts 
$\alpha_\ell(\nu)$ are, very nearly, independent of $\ell$ in the outer layers, and can all be replaced by an $\ell$ independent function of frequency $\alpha(\nu)$.

To test the validity of this approximation  we  calculate $\delta\alpha_\ell$, the difference between the  $\alpha_\ell$ and their average over values $\ell=0,1,2,3$, as a function of frequency and fractional radius $x=r/R$ for four models; the three $1.15M_\odot$ models A, B, C as defined above, and also for a solar model (Model S of Christensen-Dalsgaard et al, 1996).  All frequency sets,  except Model C, have 14 frequencies of each degree ($\ell=0,1,2,3$)  centred on a frequency  $\nu_{max} \propto M/(R^2\,T_{eff}^{1/2})$ 
 normalised to a solar value of $3050\mu$Hz.  Model C, with many mixed modes, has $14~ \ell=0$, $18 ~\ell=1$,  $18~ \ell=2$, and  $20~ \ell=3$ frequencies (see Fig 4).
For all frequency sets the maximum value of $\delta\alpha_\ell(\nu)$  almost always occurs for the lowest frequency $\ell=3$ mode, as expected on theoretical grounds
since the dominant $\ell$ dependent term in the oscillation equations $\propto \ell(\ell+1)\,c^2/(\nu^2 r^2)$.

 \begin{figure}[h]
  \vskip-6pt
  \begin{center} 
   \includegraphics[width=8.75cm]{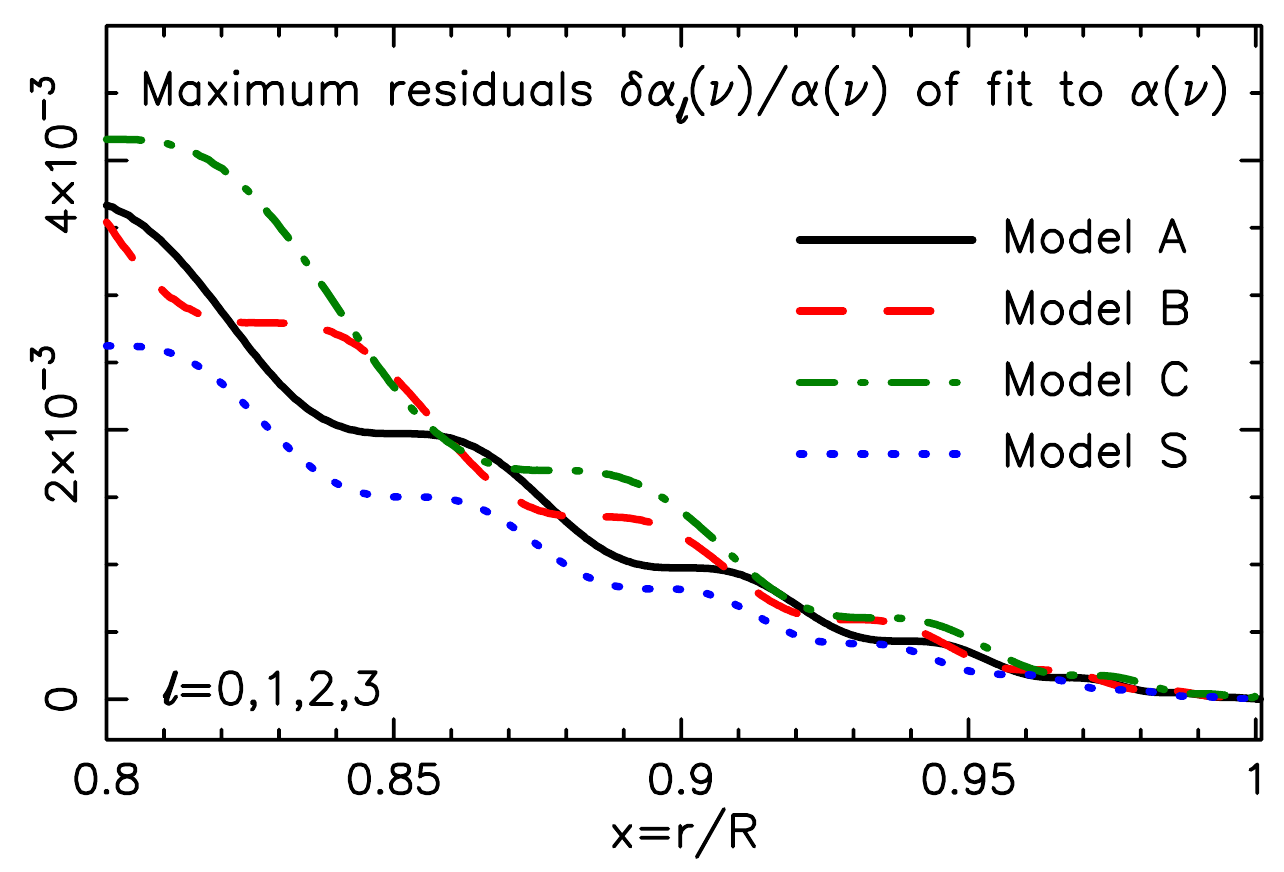}
  \end{center}  
   \vskip -14pt
   \caption{Model A: The maximum departure of the continuous $\alpha_\ell(\nu)$ from a function of frequency alone.}
   \vskip-10pt
\end{figure}

Fig. 14 shows the results - giving the maximum departure $\delta\alpha/\alpha$ as a function of fractional radius on using the continuous phase shifts, the error decreases with increasing radius but is quite large at smaller radii. Clearly we should choose a fitting point $x_f$ for  the phase matching well into the outer layers. 
\begin{figure}[h]
  \vskip-6pt
  \begin{center} 
   \includegraphics[width=8.75cm]{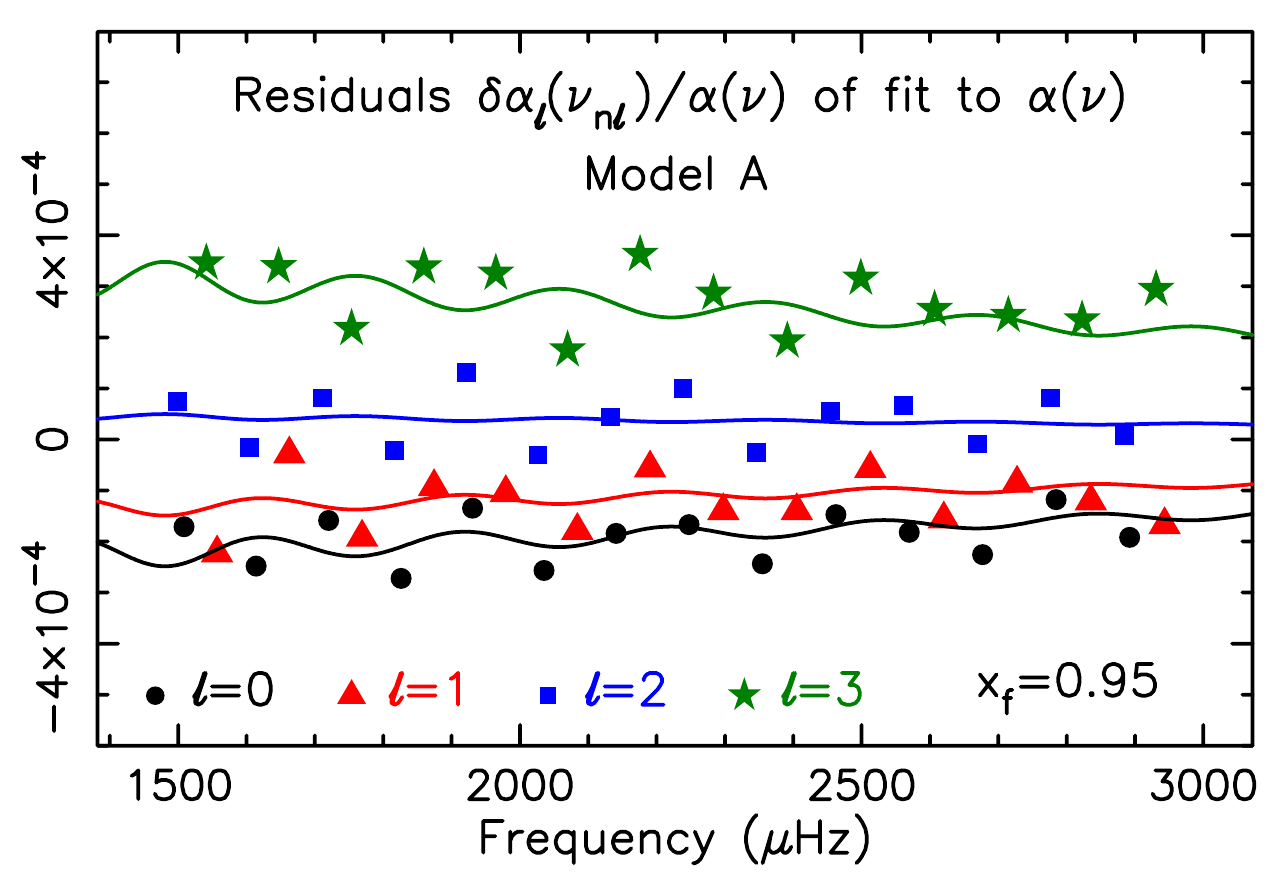}
  \end{center}  
   \vskip -14pt
   \caption{Model A: The departure of the discrete  $\alpha_\ell(\nu_{n\ell})$ from a best fit function of frequency alone compared with the errors for the continuous phases.}
   \vskip-6pt
\end{figure}

A further requirement  is that the error in fitting the phase function for the discrete values $\nu_{n\ell}$ 
with a series of Chebyshev polynomials  (or any other set of basis functions), is sufficiently small not to be important in the phase matching algorithm. One way of testing this is to compare the $\delta\alpha_\ell$ from the continuous  $\alpha_\ell(\nu)$ with those from using the discrete values.  This is shown in Fig 15 for Model A at a fitting  radius $x_f=0.95$,
where for the discrete values the average has been computed by fitting the full set  of 56 $\alpha_\ell(\nu_{n\ell})$
with a series of Chebyshev polynomials up to degree 13.  The errors are consistent with those from the continuous $\alpha_\ell(\nu)$ and
lie in the range $-2.7< \delta\alpha/\alpha < - 3.6 \,10^{-4}$, demonstrating that the requirement on the errors is satisfied.

 We then repeated the calculation for Models B, C and S  and for mode  sets of degree 
 $\ell= \{0,1\}, \{0,1,2\}, \{0,1,2,3\}$. The fractional errors are given in Table 5, along with the lowest values of $\nu$ in the full frequency set.  All the errors are small but  increase with decreasing frequency, and with increasing $\ell$.

\begin{table} [h]
\setlength{\tabcolsep}{8pt}
\centerline   
{\bf Table 5. Range of errors {\boldmath $\delta\alpha_\ell/\alpha(\nu) $ at $x_f=0.95$ - units $10^{-4}$}}
\vskip 0.15cm
\tiny
\centering
 \begin{tabular}{l c c c c c c c c c c c c c c   } 
\hline\hline 
\noalign{\smallskip}
\tiny
 $\ell$ values&Model S   & Model A& Model B&Model C \\	[0.5ex]
$\nu_{min}$ &   2218  &   1499   &  1121  & 566\\  [0.8ex]
\hline
\noalign{\smallskip}
\tiny
 0,1 & -0.4 : 0.5&-1.0  : 1.1&-1.2 :  1.3& -3.2 : 2.7 \\[0.2ex]
 0,1,2 & -1.2 : 1.2&-1.7 -: 2.1&-2.0 : 2.1& -4.1 : 4.1 \\[0.2ex]
 0,1,2,3 & -1.8 : 2.8&-2.7 : 3.6&-3.1 : 4.3&-5.8 : 5.8 \\[0.2ex]
\hline
\end{tabular}
 \end{table}
\normalsize

We also checked the accuracy with which the continuous phase shifts, computed using the surface boundary condition that the oscillating potential and its derivative are zero at the surface, reproduce the exact values for the eigenfrequencies $\nu_{n\ell}$;  the fractional error is absolutely negligible: zero for $\ell=0,1$ and   $< 10^{-6}$  for $\ell=2,3$ for all 4 models.   We also checked the error in using the Cowling approximation  (neglect of the oscillating gravitational potential); here the error is of course zero for $\ell=0$, but  is still less than  $10^{-6}$ for $\ell=1,2,3$ for all 4 models.

The second criterion for the choice of the fitting point $x_f$ is that the $\delta$s should not be influenced by the structure of the outer layers and in particular the HeII ionisation layer,
since they then have a short period quasi periodic modulation of large amplitude, the choice of $x_f=0.95$ satisfies this criterion,  as do other values in this region.

Although we calculate the continuous inner phase shifts by imposing the boundary condition on the gravitational potential at the surface, we also checked the error in imposing this condition at $x_f=0.95$, the error is $\sim 10^{-8}$. 

The error in taking $\alpha_\ell(\nu)=\alpha(\nu)$ limits the accuracy with which the phase function $\cal{G}(\ell,\nu)$ can collapse to a function only of frequency. 
For model A with  $0.5 < \delta_\ell <1.5$  and $2 < \alpha < 4$ (see Fig 2),  the $\ell$ dependent  contribution to  ${\cal G}(\ell,\nu)$ from the $\delta$s is of order $1/\pi \sim 0.3$, whereas the $\ell$ dependence of $\alpha / \pi$ is  of order $4 \, 10^{-4}$, much smaller than the contribution from the $\delta$s.

For real data, with errors on the frequencies $\sigma\sim  0.2\mu$Hz, and a large separation $\Delta\sim 100\mu$ Hz, the dominant error in $\cal{G}$ comes from $\sigma/\Delta \sim 2\,10^{-3}$,  larger than the errors from $\delta\alpha_\ell$. We conclude that the approximation $\alpha_\ell(\nu) = \alpha(\nu)$ in the outer layers is valid. 

The one exception is the Sun;  with the very high precision achieved by space and ground based experiments  it is necessary to take the outer phase shifts to be represented by two functions $\alpha_\ell(\nu) = \alpha(\nu)+\ell(\ell+1)\,\alpha_2(\nu)$  (cf Roxburgh \& Vorontsov 1994b).

\newpage

\section*{Acknowledgements}
The author thanks Sergei Vorontsov for many valuable contributions to work on phase shift analysis over two decades, and Andrea  Miglio for kindly providing his model set.  He also  gratefully acknowledges support from the Leverhulme Foundation under grant EM-2012-035/4.


\end{document}